\documentclass[pof, aip, reprint]{revtex4-1}

\usepackage{amsmath}    
\usepackage{graphicx}   
\usepackage{verbatim}   
\usepackage{color}      
\usepackage{hyperref}   

\newcommand{\DEFN}{\equiv}

\newcommand{\AV}[1]{\overline{#1}}

\newcommand{\OD}[2]{\frac{\mathrm{d}#1}{\mathrm{d}#2}}

\newcommand{\D}{\mathrm{d}}

\renewcommand{\d}{\textnormal{d}}
\newcommand{\Ri}{\textnormal{Ri}}
\renewcommand{\Pr}{\textnormal{Pr}}
\renewcommand{\Re}{\textnormal{Re}}
\begin{document}

\title{Turbulent transport and entrainment in jets and plumes: a DNS study}
\author{Maarten van Reeuwijk}
\affiliation{
Department of Civil and Environmental Engineering, Imperial College London, London SW7 2AZ, UK}

\author{Pietro Salizzoni}
\affiliation{Laboratoire de M\'{e}canique des Fluides et d'Acoustique, University of Lyon, CNRS UMR 5509 Ecole Centrale de Lyon, INSA Lyon, Universit\'e Claude Bernard, 36, avenue Guy de Collongue, 69134 Ecully, France}

\author{Gary R. Hunt}
\affiliation{Department of Engineering, University of Cambridge, Cambridge CB2 1PZ, UK}

\author{John Craske}
\affiliation{
Department of Civil and Environmental
  Engineering, Imperial College London, London SW7 2AZ, UK}

\date{\today}

\begin{abstract}

We present a new DNS data set for a statistically axisymmetric turbulent jet, plume and forced plume in a domain of size $40 r_0 \times 40 r_0 \times 60 r_0$, where $r_0$ is the source diameter.
The data set supports the validity of the Priestley and Ball entrainment model in unstratified environments (excluding the region near the source), which is corroborated further by the Wang and Law and Ezzamel \emph{et al.} experimental data sets, the latter being corrected for a small but influential co-flow that affected the statistics.
We show that the second-order turbulence statistics in the core region of the jet and the plume are practically indistinguishable, although there are significant differences near the plume edge.
The DNS data indicates that the turbulent Prandtl number is about 0.7 for both jets and plumes. 
For plumes, this value is a result of the difference in the ratio of the radial turbulent transport of radial momentum and buoyancy.
For jets however, the value originates from a different spread of the buoyancy and velocity profiles, in spite of the fact that the ratio of radial turbulent transport terms is approximately unity.
The DNS data does not show any evidence of similarity drift associated with gradual variations in the ratio of buoyancy profile to velocity profile widths.
\end{abstract}
\maketitle

\section{Introduction}
\label{sec:intro}

The mixing of buoyant fluid releases with the surrounding fluid is of
primary concern for a wide number of industrial and environmental turbulent
flows, spanning the ascending motions of thermals in the atmosphere, the
rise and fall of volcanic eruption columns, the release of
airborne pollutants or the propagation of smoke in free or enclosed spaces
\cite{Hunt1991}. Much attention has therefore been paid to the turbulence
dynamics of buoyant releases in a multiplicity of flow configurations. One of the most studied flows among these, commonly referred to as a `plume', is the
free-shear flow arising from a localised source of buoyancy. Since the
pioneering work of \citet{Zeldovich1937}, \citet{Priestley1955} and
\citet{Morton1956}, plumes have been the object of several theoretical
\cite{Morton1959}, experimental \cite{Papanicolaou1988, Panchapakesan1993,
  Shabbir1994, Wang2002} and numerical \cite{Pham2007,Plourde2008}
investigations and are well documented in a number of review articles
\citep{List1982,Woods2010,Hunt2011}. In this context, the well-known turbulent
jet can be regarded as a plume without buoyancy and provides a
reference state for understanding how buoyancy modifies the behaviour of these
free-shear flows.

Jets and plumes are canonical examples of flows that evolve in a self-similar
fashion \cite{Hunt2011}: sufficiently far from the source, a rescaling of the
radial coordinate and dependent variables by a characteristic local width $r_m$, velocity $w_m$ and
buoyancy $b_m$, results in a collapse of the data onto a single curve. The velocity and buoyancy profiles are well represented by a Gaussian form
\cite{List1982}, and self-similarity allows power laws, relating the scales $r_m$, $w_m$ and $b_m$ to the streamwise (vertical direction opposing the gravitational vector) $z$-coordinate
\citep{Morton1956}, to be deduced. 
Due to the presence of buoyancy, the $z$-dependence of plumes is markedly different to that of jets, yet in other respects, as discussed in this paper, these flows are broadly alike.

There are several ways to determine the characteristic scales $r_m$, $w_m$
and $b_m$. A popular experimental method is to capitalise on the Gaussian
shape of the velocity and buoyancy profiles, and associate $r_m$ with the standard deviation of the
Gaussian, and $w_m$ and $b_m$ with the maximum velocity and buoyancy,
respectively.  A method that does not rely directly on the assumption of a Gaussian shape is to determine local scales based on integral quantities of the flow:
\begin{equation}
r_{m} \DEFN \frac{Q}{M^{1/2}},\ \ \
w_{m} \DEFN \frac{M}{Q},\ \ \
b_{m} \DEFN \frac{B}{r_m^2},
\label{eq:scales}
\end{equation}
where the integral volume flux $Q$, specific momentum flux $M$ and buoyancy $B$ are defined as
\begin{equation}
Q \DEFN 2\int_{0}^{\infty}\AV{w}r\D r,\ \ \   
M \DEFN 2\int_{0}^{\infty}\AV{w}^{2}r\D r,\ \ \ 
B \DEFN 2\int_{0}^{\infty}\AV{b} r\D r.
\label{eq:depvar}
\end{equation}
\noindent 
Here $\overline{w}$ is the average (ensemble or time) streamwise velocity, $b=g(\rho_e - \rho)/\rho_e$ is the fluid buoyancy and $\overline{b}$ its average value, $g$ is the modulus of the gravitational acceleration and $\rho_e$ the density of the environment.
Here, $Q$, $M$ and $B$ are scaled, rather than actual, integral fluxes due to a factor $\pi$ that is not present in their definitions; this is common practice as it simplifies the resulting analytical expressions\citep{Hunt2005a}.

It should be noted that the definition of $b_m$, in \eqref{eq:scales}-\eqref{eq:depvar}, is non-standard as it is usually expressed in terms of the buoyancy flux
\begin{equation}
F \DEFN 2\int_{0}^{\infty}\AV{w}\AV{b} r\D r,
\end{equation}
as $b_m = F/Q = F/(w_m r_m^2)$.
Whilst this is a perfectly reasonable definition, it implicitly assumes averaging over a radius associated with the buoyancy profile which, in general, will not be exactly equal to $r_m$.
With a single lengthscale $r_m$ as defined in \eqref{eq:scales}, it follows that $F=\theta_m w_m r_m^2 b_m$ where $\theta_m$ is a dimensionless profile coefficient (see also section \ref{sec:profilecoefs}); thus the definition of $b_m$ in terms of $F$, in the current framework, is  $b_m = F/(\theta_m Q)$.
The profile coefficient $\theta_m$ is intimately related to the ratio of the widths of the buoyancy and velocity profiles (see section \ref{sec:profilecoefs}), plays an important role in longitudinal mixing in jets \cite{Craske2015b} and is purportedly responsible for the large scatter in measurements of plume entrainment \cite{Kaminski2005}.

The dilution of jets and plumes can be quantified by integrating the
continuity equation over the radial direction, which results in
\begin{equation}
\label{eq:Q}
\frac{1}{r_m}\OD{Q}{\zeta} = - 2 \left[ ru \right]_{\infty}.
\end{equation}
Here $\zeta \equiv \int_0^z r_m^{-1} \d z'$ is a dimensionless vertical coordinate and $\left[ ru \right]_{\infty}$ is a net volume flux into the jet/plume per unit
height. 
The entrainment assumption
\cite{TayGrep1945a, Batchelor1954, Morton1956, Turner1986}, links the radial volume flux to internal jet/plume properties via
\begin{equation}
 - \left[ ru \right]_{\infty} = \alpha r_m w_m,
\label{eq:ent}
\end{equation}
where $\alpha$ is the entrainment coefficient. Substitution of \eqref{eq:ent} into \eqref{eq:Q} and rearranging results in
\begin{equation}
\alpha = \frac{1}{2 Q} \OD{Q}{\zeta}.
\label{eq:alphadef}
\end{equation}
Thus, the entrainment coefficient can be interpreted as (half) the relative increase in volume flux over a typical jet/plume radius $r_m$. This relation also clearly establishes that $\alpha$ is a measure of dilution: the higher its value, the more fluid will be mixed into the jet/plume per (vertical) unit $r_m$.

Typical ranges of values for $\alpha$ in jets and plumes are,
respectively \citep{Carazzo2006}, $0.065<\alpha_j<0.084$ and
$0.10<\alpha_p<0.16$, which, in spite of the scatter, strongly suggests that
$\alpha_p>\alpha_j$.  
Using the observation that the spreading rates
$\d r_m / \d z$ of jets and plumes are approximately equal
\citep{List1973,List1982},
and the well-known far-field solutions $r_m=2\alpha_j z$ and $r_m = \frac{6}{5} \alpha_p z$ for jets\cite{Fischer1979} and plumes\cite{Morton1956}, respectively, it follows directly that
\begin{equation}
\alpha_p \approx \frac{5}{3} \alpha_j.
\end{equation}
By applying the relation above to the observed range of values of $\alpha_j$, we obtain $0.108<5\alpha_j/3<0.133$,
which is in reasonably good agreement with the available data for $\alpha_p$.

The fact that the spreading rates of jets and plumes are practically identical
is intimately linked with the turbulence production in the interior.
Indeed, by considering balance equations for the kinetic energy of the mean flow in jets and plumes \citep{FoxD_jgr_70a, Kaminski2005, Ezzamel2015, vanReeuwijk2015},
the spreading rate can be directly linked to the turbulence production inside
the plume. For a self-similar Gaussian plume, ignoring turbulence and pressure
effects and assuming $\theta_m=1$, it follows that \citep{vanReeuwijk2015}
\begin{equation}
\label{eq:drmdzMSG}
\OD{r_m}{z} = - \frac{3}{4} \delta_m,
\end{equation}
where
\begin{equation}
  \delta_m = \frac{4}{w_m^3 r_m} \int_0^\infty \overline{u'w'} \OD{\overline{w}}{r} r \d r
\end{equation}
is a dimensionless profile coefficient associated with the integral of
turbulence production due to shear. This quantity is generally negative as it
signifies the energy transfer from the mean to the turbulence.  Hence, under
the realistic assumptions leading to \eqref{eq:drmdzMSG}, it follows that
$\delta_m$ is solely responsible for the plume spread, and identical spreading
rates imply identical values for $\delta_m$.  Direct estimations, either using
flow measurements or with high-fidelity simulations, confirm that the value of
$\delta_m$ for jets and plumes is indeed nearly identical
\citep{vanReeuwijk2015}. 

Using the equation for mean kinetic energy, it is possible to derive
entrainment relations that fundamentally link $\alpha$ to the production of
turbulence kinetic energy, the Richardson number and shape effects.  For a self-similar Gaussian plume with $\theta_m=1$, ignoring turbulence and pressure
effects\cite{FoxD_jgr_70a}, the entrainment relation is
\cite{vanReeuwijk2015}
\begin{equation}
\label{eq:alphaMSG}
\alpha = -\frac{3}{8} \delta_m + \frac{1}{4} \Ri,
\end{equation}
where the Richardson number $\Ri$, defined as
\begin{equation}
\label{eq:Ri}
\Ri = \frac{b_m r_m}{w_m^2} = \frac{BQ}{M^{3/2}}
\end{equation}
characterises the significance of buoyancy compared with inertia.
An important implication of the fact that $\delta_m$ does not differ between jets and plumes (i.e.\ is constant) is that \eqref{eq:alphaMSG} shows that the difference in $\alpha$ is caused purely by the influence of mean buoyancy via $\Ri$.  
By using the observation that $\delta_m$ is a constant, \eqref{eq:alphaMSG} can be rewritten as \citep{vanReeuwijk2015}
\begin{equation}
\label{eq:alphaPB}
\alpha = \alpha_j + (\alpha_p - \alpha_j)\Gamma
\end{equation}
which is commonly referred to as the Priestley and Ball entrainment
model \citep{FoxD_jgr_70a, Priestley1955}.  Here, $\Gamma=\Ri/\Ri_p$ is the
flux balance parameter, where $\Ri_p = 8 \alpha_p \beta_g/5$  is the Richardson number for a pure plume\citep{vanReeuwijk2015} and $\beta_g$ is a profile coefficient associated with the total momentum flux (see section \ref{sec:profilecoefs} for its definition).
The condition $\Gamma=1$ represents a stable equilibrium (with respect to perturbations in $\Gamma$), a condition referred to as that of a `pure plume'. 
The other equilibrium condition is given by $\Gamma=0$, i.e. that of a `pure jet', a condition which is however unstable to the addition of an arbitrarily small amount of buoyancy \cite{Hunt2005a}. 
For forced plumes, which have an excess of momentum (relative to pure plume conditions) at the source\cite{Morton1959}, $ 0 < \Gamma < 1$ , whereas $\Gamma > 1$ for lazy plumes, which have a deficit of momentum \cite{Hunt2005a}.  
Previous experimental studies observed that \eqref{eq:alphaPB} accurately describes the behaviour of jets, plumes and forced plumes \citep{Wang2002, Ezzamel2015}.

If the magnitude of the dimensionless turbulence production $\delta_{m}$ is
approximately equal in jets and plumes, one is led to ask what this implies about the radial transport of scalar quantities in the flow. The
turbulent Prandtl number

\begin{equation}
\Pr_T= \frac{\nu_T}{D_T},
\label{eq:PrT}
\end{equation}
where $\nu_T$ and $D_T$ are the eddy viscosity and eddy diffusivity, respectively,
quantifies the effectiveness with which the flow mixes momentum compared with
buoyancy/mass and is a useful quantity in this regard. The consensus is that
$\Pr_T=0.7$ in axisymmetric jets and plumes \cite{Chen1980}, which suggests that turbulence transports buoyancy/mass more efficiently than momentum
\cite{Chua1990} in both cases. 
However, the underlying physics and their implications for entrainment and for the relative widths of the scalar profile compared with the velocity profile are not understood. 
For jets there is a good agreement between investigators that suggests the scalar field is wider than the velocity field \cite{Fischer1979, Chen1980, Papanicolaou1988, Wang2002}. 
For plumes however, as discussed in \cite{Hunt2001} and elsewhere, there is significant uncertainty: some studies reveal that the velocity field is wider than the buoyancy field \cite{Shabbir1994, George1977, Chen1980}, others reveal that it is narrower \cite{Ezzamel2015, Papanicolaou1988, Fischer1979, Rouse1952}; several results imply that the velocity and scalar profiles have roughly the same width \cite{Wang2002, Devenish2010} and some imply that the relative width varies with height \cite{Kaminski2005}. 
The present paper seeks to untangle the confusion regarding the relationship between $\Pr_T$ and the widths of the scalar and velocity profiles by supplementing the available experimental data with precise information from direct numerical simulation (DNS).

Herein, we follow the approach of \citet{Ezzamel2015} by performing a side-by-side comparison of turbulent jets, plumes and the intermediate case of a forced plume, but using DNS rather than laboratory experiments.
With DNS it is relatively straightforward to prescribe boundary conditions consistent with the analytical solutions and furthermore, DNS provides access to all variables, including pressure, at Kolmogorov-scale resolutions.
In section \ref{sec:sims}, the simulation details are presented.
Integral flow statistics, such as the evolution of $\Gamma(z)$, are presented in section \ref{sec:integral} and the deduced entrainment coefficient $\alpha$ is shown to follow closely the Priestley and Ball entrainment model \eqref{eq:alphaPB}.
Self-similarity of the first- and second-order statistics is discussed in section \ref{sec:selfsim}, which includes an analysis of the invariants of the anisotropy tensor.
Profile coefficients, which represent the relative contribution of various physical processes relative to the characteristic scales are presented in section \ref{sec:profilecoefs}, and these are used to decompose the entrainment coefficient into its individual components in section \ref{sec:alpha}. 
Section \ref{sec:nuT} discusses the radial turbulent transport of streamwise momentum and buoyancy, as quantified by the eddy viscosity $\nu_T$ and diffusivity $D_T$. The turbulent Prandtl number will be decomposed and it is shown that even though jets and plumes share a very similar value for $\Pr_T$, the underlying reason in each case is different.
Concluding remarks are made in section \ref{sec:conclusions}.

\section{Simulation details}
\label{sec:sims}
We simulate axisymmetric jets and plumes driven by an isolated
source of steady specific momentum flux $M_{0}$, volume flux $Q_{0}$ and
buoyancy flux $F_{0}$. The source is approximately circular and
located at the centre of the base of a cuboidal domain of size
$40^2\times 60$ source radii, $r_{0}$. The fluid motion is
governed by the incompressible Navier-Stokes equations under the
Boussinesq approximation, which we solve numerically using
$1280^{2}\times 1920$ computational cells over a uniform Cartesian
grid. The code for the DNS employs a spatial discretisation of
fourth-order accuracy that conserves volume, momentum and
energy, and integration in time is performed using a third-order
Adams Bashforth scheme \citep[further details can be found in][]{Craske2015}. 
On the vertical and top faces of the domain we impose open boundary conditions.
These allow fluid to enter and leave the domain in a manner that is consistent with flow in an unconfined domain\citep{Craske2013}. 
We initiate the turbulence by applying uncorrelated perturbations of $1\%$ to the velocities in the first cell above the source.

To simulate the jet J we impose a constant uniform vertical
velocity $w_{0}$ at the source. Consequently, a constant scalar
flux can be maintained by imposing a Dirichlet boundary condition
$b=b_{0}$ on a given scalar quantity $b$ at the source. 
For the jet simulation J, this scalar quantity is passive, i.e. its presence does not imply a source term in the momentum equation.
In the forced plume simulation F, for which $b$ corresponds to buoyancy, the Dirichlet boundary condition on $b$ at the source
results in a positive buoyancy flux $F_{0}$. 
The source conditions used in the simulation of plume P correspond to $w_{0}=0$ and a specified positive integral buoyancy flux $F_{0}$; in practice, the buoyancy flux $F_{0}$ is a diffusive flux resulting from a Neumann condition on the buoyancy at the source. 
Therefore, the plume simulation P is infinitely lazy at the source ($\Gamma_{0} \equiv 5F_{0}Q_{0}^{2}/(8\alpha_{p} M_{0}^{5/2})=\infty$) although, over a relatively short distance, plume P becomes pure.
Based on the analysis of \citet{Hunt2005a}, in which a constant entrainment coefficient model is assumed, the rate of decrease of the local Richardson number immediately above a highly-lazy plume source scales as 
\begin{equation} \frac{d \Gamma }{d \zeta }\bigg|_{\zeta=0} \propto - \Gamma_0^{9/5}.
\end{equation}
Thus, the vertical distance required to approach pure-plume behaviour reduces to zero as the laziness of the source increases, i.e. as
$\Gamma_0 \rightarrow \infty$. As a consequence, our plume arising from the heated disc boundary condition, which represents the limit of an infinitely lazy plume source, is expected to establish pure-plume behaviour immediately above the source and, as such, to closely mimic a true pure-plume source.
For jet J and forced plume F we define the
source Reynolds number $\Re_{0}\equiv 2 M_{0}^{1/2}/\nu$ and for
plume P, $\Re_{0}\equiv 2 F_{0}^{1/3}r_{0}^{2/3}/\nu$. 
The calculated values of $\Re_{0}$, in addition to further details of the simulations, can be found in Table \ref{tab:sim}.

\begin{table}
\begin{center}
\def~{\hphantom{0}}
\begin{tabular*}{\textwidth}{l @{\extracolsep{\fill}} ccccc|cc|ccc}
& $N_{x}N_{y}N_{z}$
& $L_{x}L_{y}L_{z}/r_{0}$
& $\textnormal{Re}_{0}$
& $\Gamma_{0}$
& $t_{\mathrm{run}}/\tau_{0}$ 
& $\alpha$ 
& $z_v/r_0$
& $a_w$ 
& $a_b$ 
& $\langle \Pr_T \rangle$ \\ 
\hline
J\ \ & $1280^{2}\times 1920$ & $40^{2}\times 60$ & 5000 & 0 & 400 &    0.067 &   -3.66 &  0.12 &    0.14 &   0.72\\ 
F & $1280^{2}\times 1920$ & $40^{2}\times 60$ & 5000 & $\approx 0.03$ & 480 & varies & varies & & & \\ 
P & $1280^{2}\times 1920$ & $40^{2}\times 60$ & 1667 & $\infty$ & 480 & 0.105 &   -3.90 &  0.13 &    0.15 &   0.68 
\end{tabular*}
\end{center}
\caption{
Simulation details. The entrainment coefficient $\alpha$ and virtual origin $z_v$ are determined directly from $r_m$ (see Fig.\ \ref{fig:scaling}). 
The constants $a_w$ and $a_b$ are prefactors of the mixing lengths of velocity and buoyancy, respectively (Eq.\ \ref{eq:lm}). $\langle \Pr_T \rangle$ is the typical turbulent Prandtl number \eqref{eq:PrT}.}
\label{tab:sim}
\end{table}

Statistics were acquired from each simulation over a duration that
is large in comparison with the typical turnover time.
For jet J and forced plume F, the turnover time based on the source conditions is $\tau_{0}\equiv r_{0}^{2}/M_{0}^{1/2}$. For plume P, $\tau_{0}\equiv
r_{0}^{4/3}/F_{0}^{1/3}$. 
Prior to obtaining
statistics we ensure that transient effects arising from initial
conditions are imperceptible in the leading-order statistics. Statistics were gathered over the time-period shown in Table \ref{tab:sim}.

Azimuthally averaged data was obtained by partitioning the domain
into concentric cylindrical cells and averaging over all cells
lying within a given shell. To compute integrals over lateral
slices of the jet (for the definition of these integrals see
section \ref{sec:profilecoefs}), we define the upper limit of integration $r_{d}$ according to $\overline{w}(r_d,z,t) = 0.02 \,\overline{w}(0,z,t)$. 

Detailed validation of the jet and plume simulations was performed in previous work \cite{Craske2015a, Craske2016} for simulations at identical $\Re_0$.
The results presented below are for a larger domain and are obtained with even higher resolutions.
A detailed validation will thus not be repeated here; agreement with existing data will be pointed out in the text and, where appropriate, included in the figures.

\section{Results}

\subsection{Integral flow statistics}
\label{sec:integral}

\begin{table}
\begin{center}
\begin{tabular}{c}
\end{tabular}
\def~{\hphantom{0}}
\vspace{-8mm}
\begin{align*}
  &          &    
  &\mathrm{Jet}  &
  &\mathrm{Plume} \nonumber \\
  \Gamma  & =  & 
  &              0&
  &              1 \\
   \mathrm{Ri} &=&  
  &              0&
  &8\alpha_{p}\beta_{g}/5 \\
   r_{m}     &   = &
  &2\alpha_{j}z   &
  &\dfrac{6}{5}\alpha_{p}z \\
   w_{m}    &  =   &
  & \dfrac{M_0^{1/2}}{2\alpha_j} z^{-1}&
  & \dfrac{5}{6\alpha_p}\left(\dfrac{9}{10}\dfrac{\alpha_p }{\theta_m\beta_g}F_0\right)^{1/3} z^{-1/3} \\
  b_{m}    &    =  &
  &\dfrac{F_0}{2 \alpha_j \theta_m } M_0^{-1/2} z^{-1} &
  & \dfrac{5 F_0}{6 \alpha_p\theta_m} \left(\dfrac{9}{10}\dfrac{\alpha_p}{
                                        \beta_g \theta_m} F_0\right)^{-1/3}
                                        z^{-5/3}
\end{align*}
\end{center}
\caption{ Asymptotic far-field solutions of jets and plumes including
  turbulence and pressure effects. In the expressions above, $M_0$ and $F_0$ are the mean specific momentum and buoyancy fluxes far away from the source.}
\label{tab:jetplumediagram}
\end{table}

\begin{figure}
\centering
\includegraphics{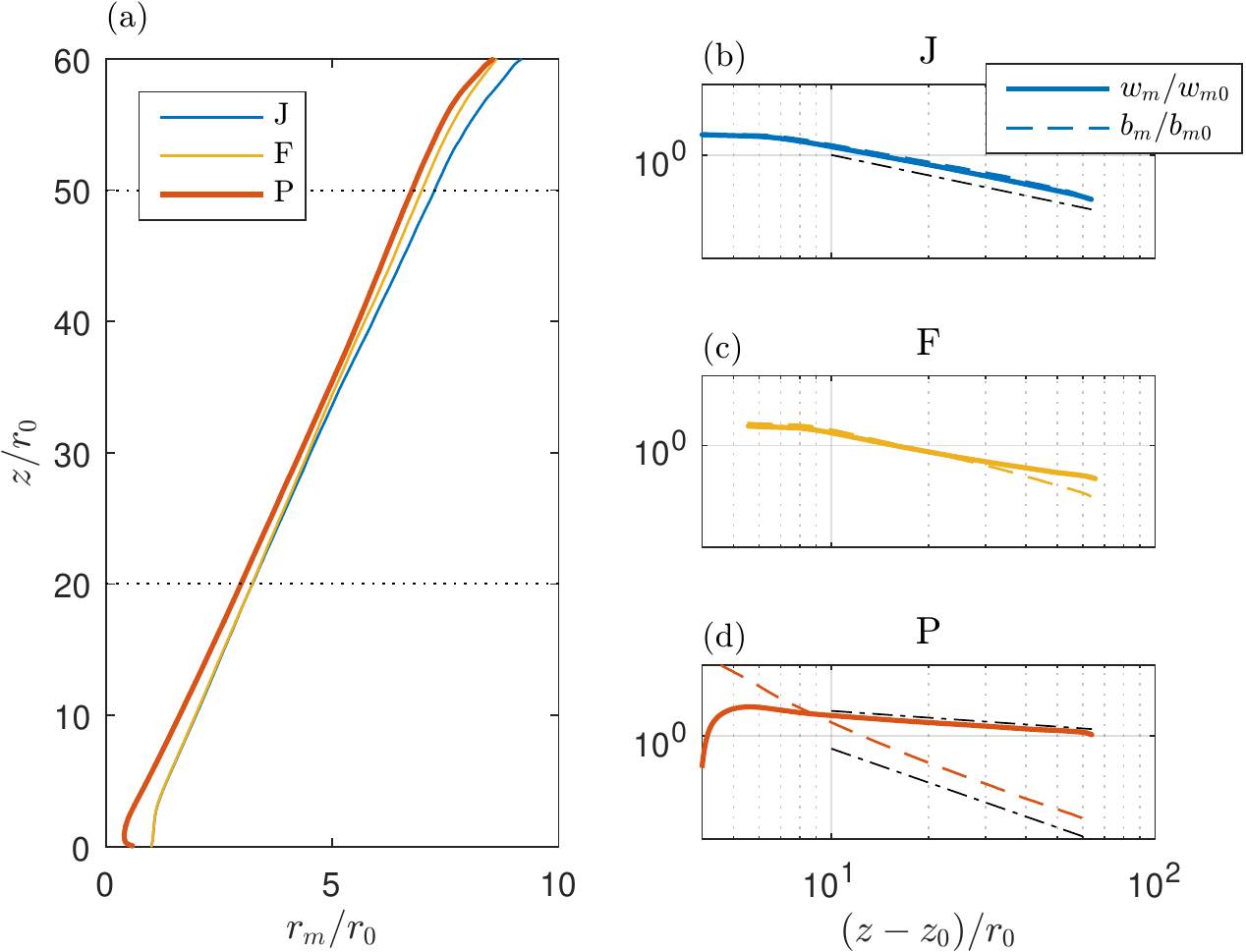}
\caption{Variation of the characteristic plume quantities with height $z$ for simulations J, F and P. (a) $r_m(z)$. (b) $b_m$, $w_m$ for J release. (c) $b_m$, $w_m$ for F release. (d) $b_m$, $w_m$ for P release. Dash-dotted lines in Figs (b-d): asymptotic power-law scaling (Table\ \ref{tab:jetplumediagram}).}
\label{fig:scaling}
\end{figure}

From an integral perspective, the plume dynamics are fully determined by the evolution of the characteristic radius $r_m$, velocity $w_m$ and buoyancy $b_m$.
For the limiting cases of a pure jet ($\Gamma=0$) and of a pure plume ($\Gamma=1$), the scaling of these parameters with the distance from the source takes the form of a power law, which can be derived from the plume equations \citep{Morton1956}.
Recently\cite{vanReeuwijk2015}, these solutions were extended to account for turbulence and pressure effects via the profile coefficient $\beta_g$ and for differences in the widths of velocity and buoyancy profiles via the coefficient $\theta_m$ (Table\ \ref{tab:jetplumediagram}). 
The profile coefficients $\beta_g$ and $\theta_m$ will be defined rigorously in section \ref{sec:profilecoefs}.
The streamwise evolution of $r_m$ is shown in Fig.\ \ref{fig:scaling}(a), confirming the almost identical linear spreading rate for the three simulations considered.
Figs \ref{fig:scaling}(b-d) show that the jet and plume both exhibit the expected power-law scaling.
The forced plume transitions from a near-field jet-like scaling   to a far-field plume-like scaling. 

As visible in Fig.\ \ref{fig:scaling}(a), the outflow boundary condition appears to affect the statistics in the upper part of the domain.
This is caused by subtle modification of the  mean flow near the outflow boundary, presumably because of slight pressure gradients \citep{Craske2013}.
These small disturbances affect the integral quantities $Q$, $M$ and $F$ via the thresholding technique (which is based on $\overline{w}$, see section \ref{sec:sims}).
Throughout what follows, all considerations on the dynamics of the flow will therefore be based on the analysis of the flow statistics for $z/r_0 <50 $.

For the two limiting cases J and P, the plume radius $r_m(z)$ is fitted in the far field ($20<z/r_m < 50$) to the analytical solutions $r_m = a \alpha (z-z_v)$, where $z_v$ is the virtual origin \cite{Hunt2001} and $a=2$ for jets and $a=6/5$ for plumes (see Table\ \ref{tab:jetplumediagram}).
We obtain $\alpha_j=0.067$ and $\alpha_p=0.105$, values that agree well with the literature and provide evidence of enhanced dilution within a plume compared to a jet.

\begin{figure}
\centering
\includegraphics{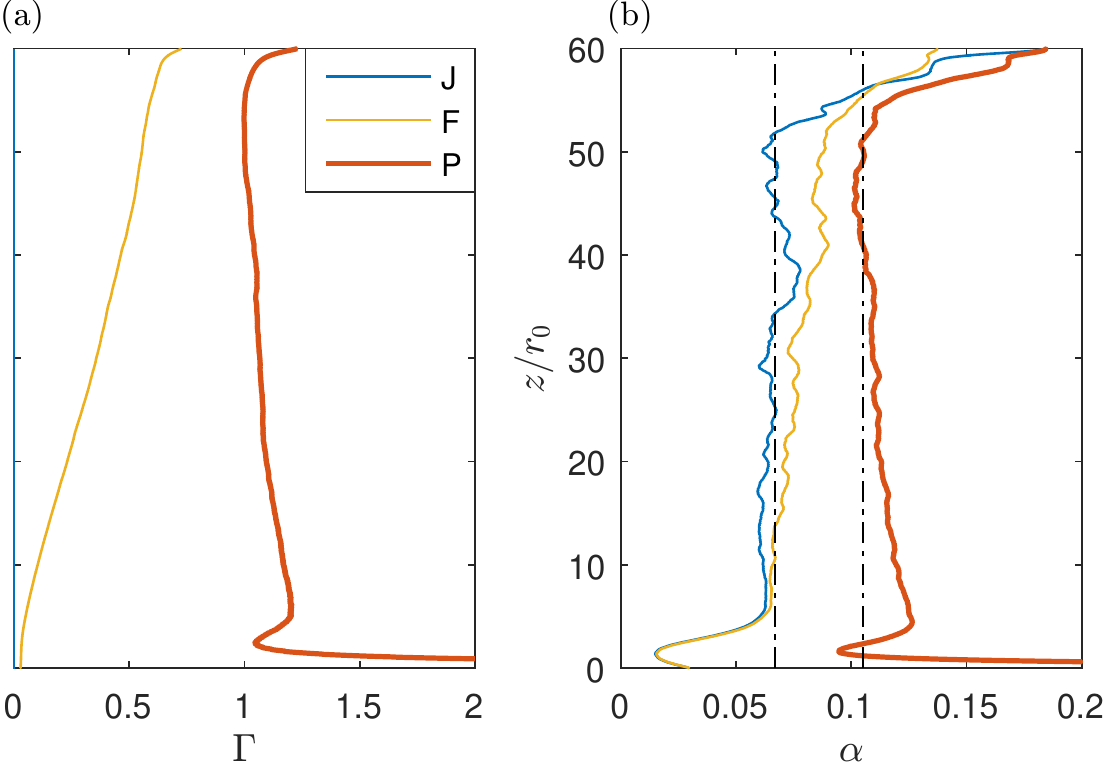}
\caption{Vertical evolution for the simulations J, F and P of: (a) the flux balance parameter $\Gamma$, and (b) the entrainment coefficient $\alpha$ computed from (\ref{eq:alphadef}).}
\label{fig:Gammaalpha}
\end{figure}

A flux balance parameter $\Gamma(z)$ that takes into account turbulence, pressure effects and differences in profile widths is defined as \citep{vanReeuwijk2015}
\begin{equation}
  \label{eq:Gamma}
  \Gamma = \frac{5 F Q^2}{8 \alpha_p \beta_g \theta_m M^{5/2}},
\end{equation}
and its variation with height is shown for the three simulations in Fig.\ \ref{fig:Gammaalpha}(a). 
For simulation J, $\Gamma$ is identically zero for all values of $z$. 
For simulation P, $\Gamma\approx 1$ except for a rapid variation in the very near field $z/r_0 < 5$. 
It is worth noting that for simulation P, the turning points of $\Gamma$ in the near field are not compatible with classic solutions of the plume equations \cite{Hunt2005a}, and have to be attributed to the near-field variations of the profile coefficients (section \ref{sec:profilecoefs}).
For forced plume simulation F, $\Gamma$ evolves approximately linearly towards its equilibrium state $\Gamma=1$, a condition which is however not attained at the upper limit of the simulated domain.

The variation of the entrainment coefficient $\alpha$ with the vertical coordinate $z$, as determined from \eqref{eq:alphadef}, is plotted in Fig.\ \ref{fig:Gammaalpha}(b).
Here, $Q$ was filtered to smooth out occasional small step changes in its value caused by the thresholding, which would otherwise result in unphysical spikes in $\d Q / \d \zeta$ and $\alpha(z)$. 
The values of $\alpha_j$ and $\alpha_p$ (Table \ref{tab:sim}) inferred from $r_m$ are displayed with the dash-dotted lines and are in good agreement with the far-field values for the jet and the plume, respectively.
The entrainment in the pure jet shows a high variability in the near field but rapidly attains the constant value $\alpha_j$, within no more than five source radii.
The entrainment coefficient for simulations J and F are almost the same in the near field. 
However, with increasing distance from the source, the entrainment coefficient in the forced plume simulation F shows a clear increasing trend. 
For the pure plume, the entrainment coefficient is very large in the near field ($z/r_0 < 5$) and then attains an approximately constant value, which is in close agreement with the far-field estimate $\alpha_p=0.105$ obtained from $r_m$. 
These results are in agreement with previous experimental investigations \cite{Wang2002, Ezzamel2015}, and show a clear tendency of the entrainment coefficient to increase with increasing $\Gamma$.

\begin{figure}
\centering
\includegraphics{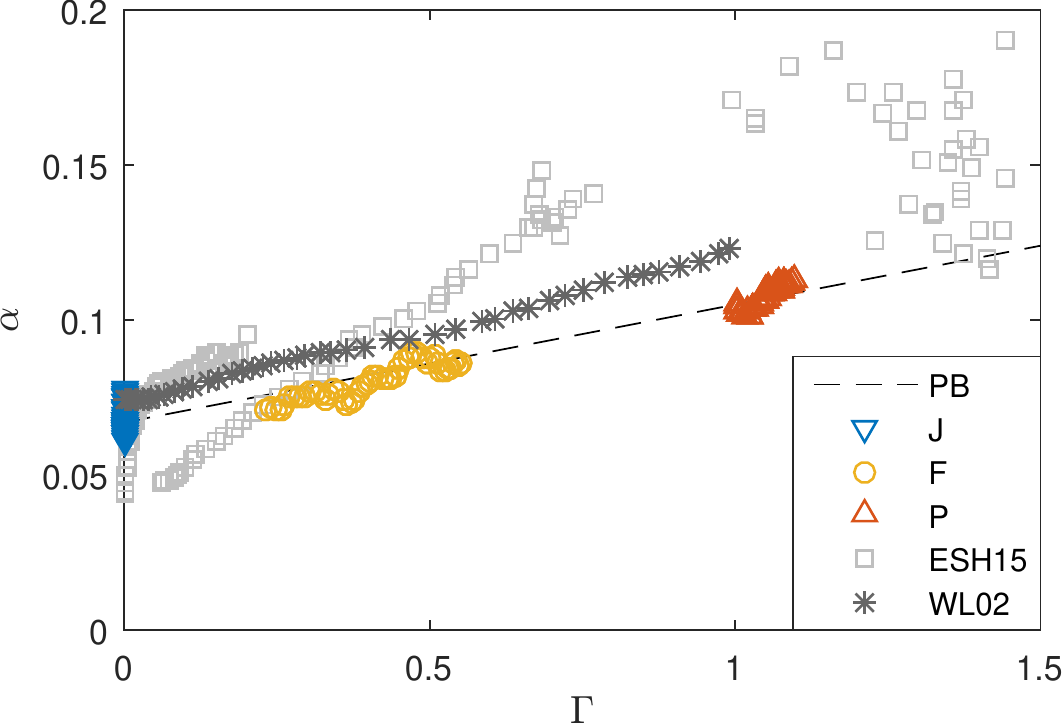}
\caption{Entrainment coefficient $\alpha$ as a function of $\Gamma$ over the interval $20<z/r_0<50$ for simulations J, F and P, confirming the good agreement with the \citet{Priestley1955} (PB) entrainment model. Data of \citet{Wang2002} (WL02) and \citet{Ezzamel2015} (ESH15) are also shown.}
\label{fig:alphaRi}
\end{figure}

By plotting the computed values of $\alpha$ as a function of $\Gamma$, it is possible to test directly the appropriateness of the \citet{Priestley1955} (PB) entrainment model (\ref{eq:alphaPB}) (cf.\ Fig.\ \ref{fig:alphaRi}).
Shown in the same plot is the experimental data from \citet{Wang2002} (WL02) and the recent measurements from \citet{Ezzamel2015} (ESH15).
The latter has been reprocessed in Appendix \ref{app:ESH15} to better represent the co-flow in the ambient which significantly influences the entrainment statistics.
The new ambient-flow correction shows much better agreement between the volume-flux based estimate of $\alpha$ and that obtained from the entrainment relation, although the data does not display the constant value of $\alpha$ that one would expect from self-similarity in the far field for the jet and plume experiments.

As is evident from Fig.\ \ref{fig:alphaRi}, all data sets show a dependence on $\Gamma$.
The current DNS data set and the WL02 data convincingly demonstrate the linear dependence on $\Gamma$ of the Priestley and Ball entrainment model \eqref{eq:alphaPB} for unstratified environments in the self-similar regime.
However, the figure also exposes the variability in what may be regarded as the limiting (or end member) entrainment coefficients; the values one would choose for $\alpha_j$ and $\alpha_p$ in \eqref{eq:alphaPB} would be slightly different for the WL02 and  current data set.
The dashed line shows the PB entrainment model using the values of $\alpha_j$ and $\alpha_p$ presented in Table \ref{tab:sim}, and good agreement with the DNS data can be observed.
The ESH15 data confirms the appropriateness of the PB model qualitatively, but despite the ambient-flow correction (Appendix \ref{app:ESH15}) the data remains noisy.
The linear dependence of $\alpha$ on $\Gamma$ implies that $\delta_m$ is practically identical in jets and plumes, as argued in the Introduction.
The entrainment coefficient will be decomposed into its various parts in section \ref{sec:alpha}.

\subsection{Self-similarity}
\label{sec:selfsim}
\begin{figure}
\centering
\includegraphics{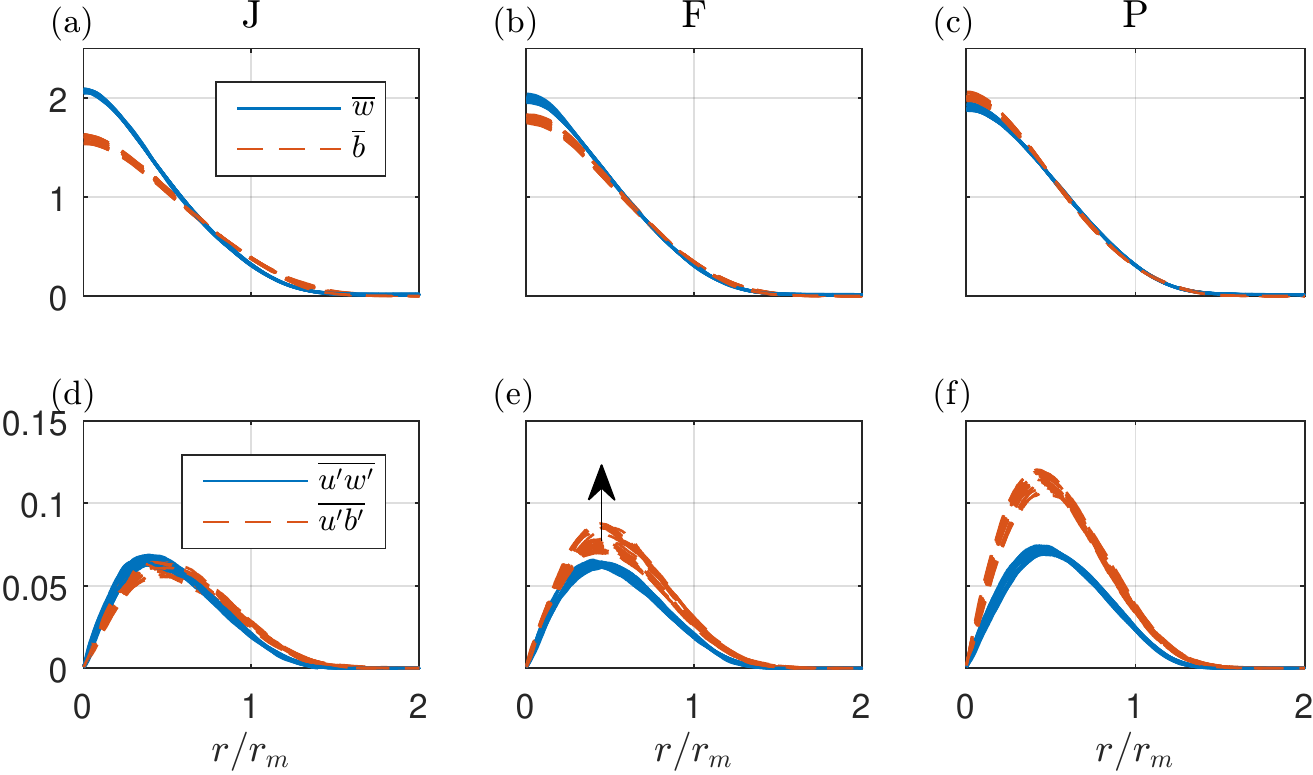}
\caption{Self-similarity profiles of $\overline{w}$, $\overline{b}$, $\overline{u'w'}$ and $\overline{u'b'}$ over the interval $20 < z/r_0<50$. (a,d): jet simulation. (b,e): forced plume simulation. (c,f): pure plume simulation.}
\label{fig:ss_first}
\end{figure}

Shown in Fig.\ \ref{fig:ss_first} are the mean velocity $\overline{w}$, buoyancy $\overline{b}$, radial turbulent momentum flux $\overline{u'w'}$ and turbulent buoyancy flux $\overline{u'b'}$ over the vertical interval $20 < z/r_0 < 50$.
As is customary, all variables are presented in dimensionless form, normalised by the local value of $r_m$, $b_m$ and $w_m$.
In line with our expectations, for all three simulations the mean vertical velocity $\overline{w}$  collapses onto a single profile which closely resembles a  Gaussian profile.

The radial profiles of mean buoyancy $\overline{b}$ also exhibit a clear Gaussian-like dependence on the radial coordinate. However, the centreline values and spread differ for the three simulations. 
Profiles for velocity and buoyancy almost coincide for plumes (Fig.\ \ref{fig:ss_first}(c)), whereas  for the forced plume and the jet, the buoyancy profiles have a slightly larger spread (as further quantified by the profile coefficient $\theta_m$ associated with mean scalar transport, see section \ref{sec:profilecoefs}).
As the integral under the dimensionless curves is unity by construction, a wider profile will reduce the centreline value of $\overline{b}/b_m$, particularly since small changes far from the centreline contribute significantly to the integral due to the conical geometry.

The profile of the turbulent radial momentum flux $\overline{u'w'}$ is practically identical for the jet, forced plume and pure plume (Figs \ref{fig:ss_first}(d-f)), which is consistent with the notion of the profile coefficient associated with the production of turbulence kinetic energy $\delta_m$ being insensitive to $\Gamma$.
However, the normalised radial turbulent buoyancy flux shows large variations in amplitude.
For the jet simulations, the profiles of $\overline{u'w'}$ and $\overline{u'b'}$ are practically identical.
For the plume simulation, $\overline{u'b'}$ is about 60\% larger in amplitude than $\overline{u'w'}$.
The profile of $\overline{u'b'}$ for the forced plume transitions smoothly from the jet profile to the plume profile as $\Gamma$ tends to unity, as indicated by the arrow in Fig.\ \ref{fig:ss_first}(e); this is in contrast to Fig.\ \ref{fig:ss_first}(f), where no systematic variation with height is present.

\begin{figure}
\centering
\includegraphics{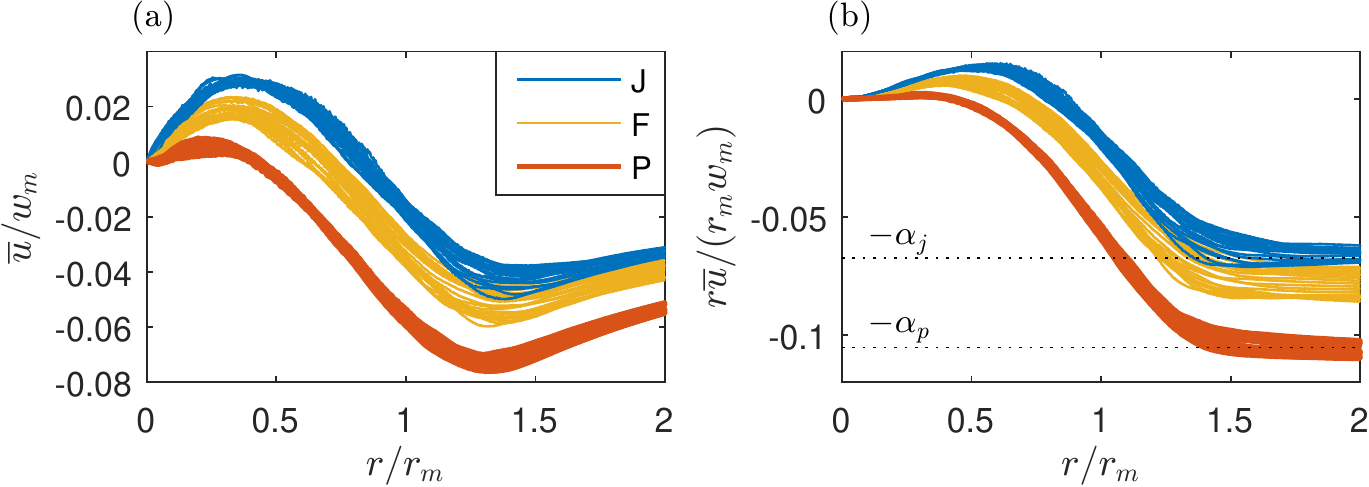}
\caption{(a) Self-similar profiles for mean radial velocity $\overline{u}$. (b) Normalised mean radial specific volume flux. The dotted lines indicates the values of $\alpha_j$ and $\alpha_p$ in Table \ref{tab:sim}.}
\label{fig:u}
\end{figure}

The normalised mean radial velocity $\overline{u}$ is shown in Fig.\ \ref{fig:u}(a). 
Contrary to the mean vertical velocity $\overline{w}$ profiles, the shape of $\overline{u}$ differs significantly between the jet, forced plume and pure plume.
For the jet, $\overline{u}$ increases from a value of zero (imposed by the radial symmetry of the flow), reaches a peak at $r/r_m \approx 0.5$, then decreases, becomes negative with a minimum at $r/r_m \approx 1.4$, after which the velocity $\overline{u}$ decays approximately inversely proportional to the radius due to the fact that the flow varies very slowly with $z$.
For the plume, the maximum in $\overline{u}$ is significantly smaller, implying a reduction in the mean outward radial transport in a plume.
The normalised specific radial volume flux $r \overline{u} / (r_m w_m)$, shown in Fig.\ \ref{fig:u}(b) for all three simulations, tends to a constant value outside the plume for $r/r_m > 1.5$.
By rearranging Eq. \eqref{eq:ent}, it clear that the constant value is equal to the entrainment coefficient $\alpha$.
The dashed lines in Fig.\ \ref{fig:u}(b) are the values of $\alpha$ in Table \ref{tab:sim} -- excellent agreement is shown with the values deduced from $r_m$.

\begin{figure}
\centering
\includegraphics{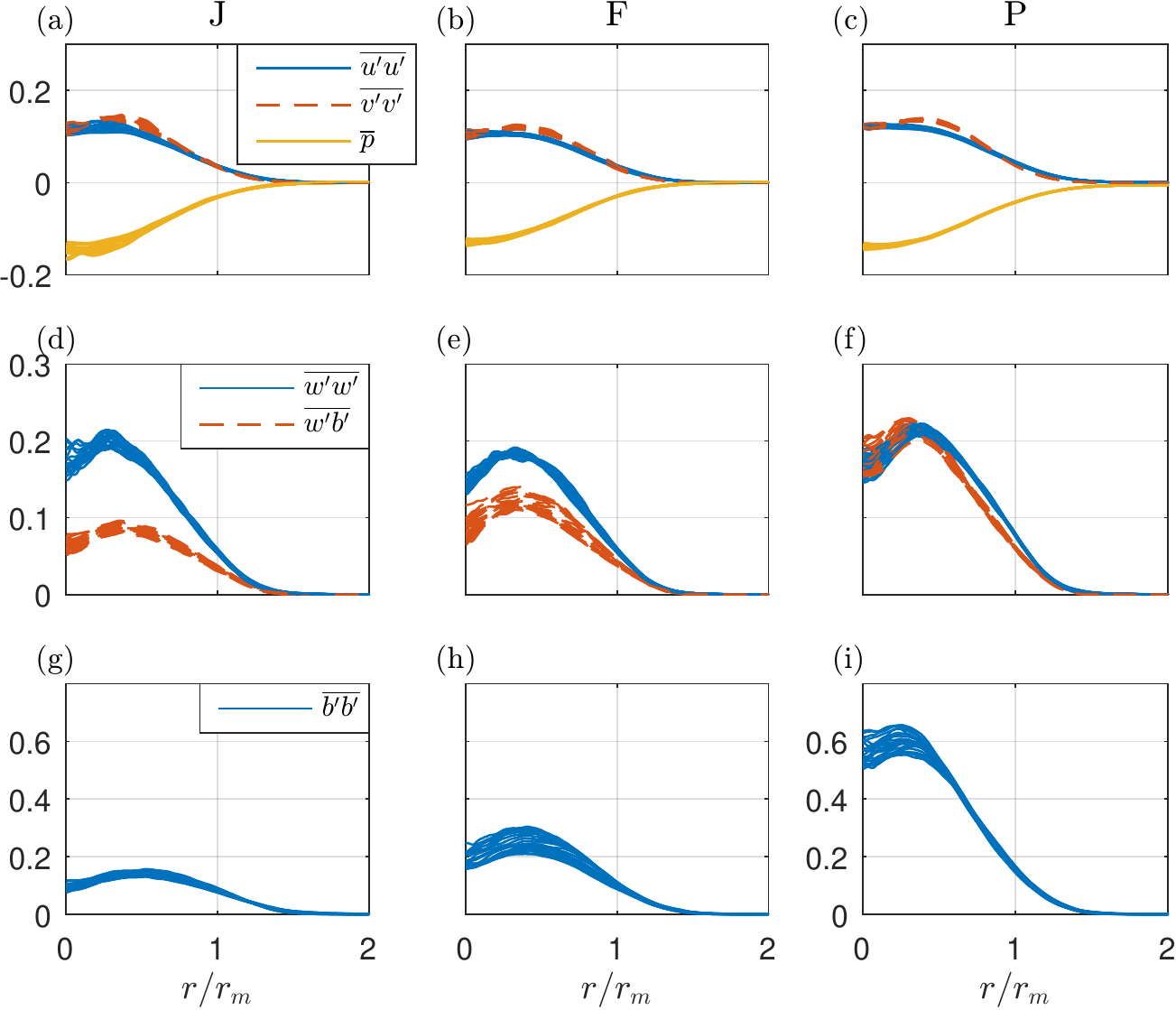}
\caption{Self-similarity profiles of second-order quantities and pressure. All quantities are normalised.}
\label{fig:ss_second}
\end{figure}

\begin{figure}
\centering
\includegraphics{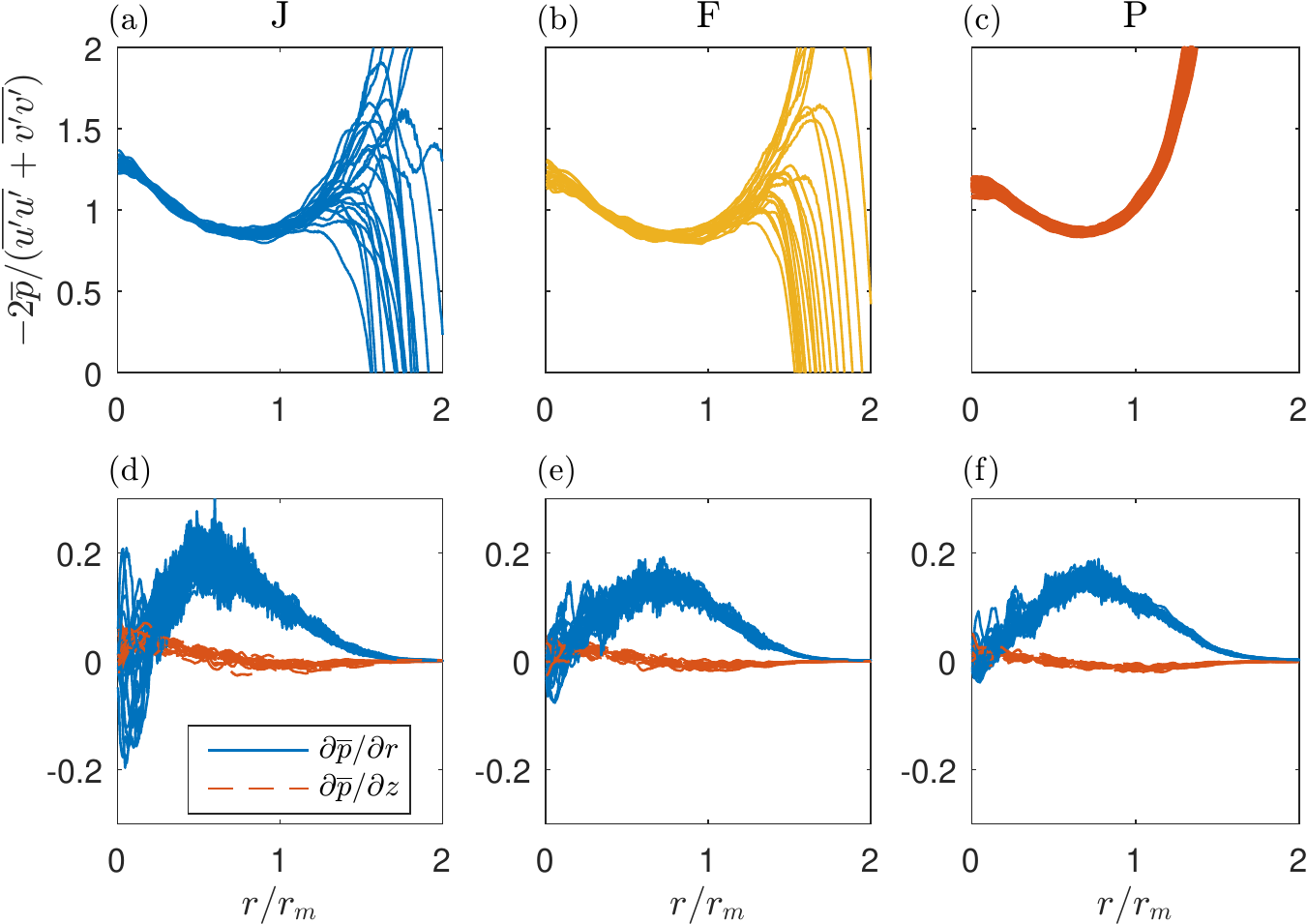}
\caption{ (a-c) The ratio of mean pressure $\overline{p}$ and $-(\overline{u'u'} + \overline{v'v'})/2$. (d-f) The horizontal and vertical pressure gradients. (a,d) Simulation J. (b,e) Simulation F. (c,f) Simulation P.}
\label{fig:pressure}
\end{figure}

The turbulent components $\overline{u'u'}$ and $\overline{v'v'}$, shown as a function of $r/r_m$ in Figs \ref{fig:ss_second}(a-c), are self-similar and practically identical.
Furthermore, their dependence on $\Gamma$ is negligible, providing further confirmation that the turbulence inside plumes and jets is similar, at least in terms of the second-order statistics.
The mean pressure $\overline{p}$ is extremely difficult to measure in laboratory experiments, and is usually approximated by \cite{Hussein1994, Wang2002,Ezzamel2015} $\AV{p} \approx -(\AV{u'u'} + \AV{v'v'})/2$.
The quantity $\overline{p}$ is readily available in DNS and it is clear from Figs \ref{fig:ss_second}(a-c) that it correlates well with $-(\AV{u'u'} + \AV{v'v'})/2$, although upon closer inspection (Fig.\ \ref{fig:pressure}) it becomes evident that $-(\AV{u'u'} + \AV{v'v'})/2$ underestimates $\overline{p}$ by  30\% in the core of the flow, whilst it overestimates $\overline{p}$ by about 10\% near $r/r_m=1$. Thus, the DNS data demonstrates that $\overline{p} = -(\AV{u'u'} + \AV{v'v'})/2$ within, say, 20\% (see \citet{Hussein1994} for a detailed explanation of the various sources of error).
Like the gradient of all quantities in a slender turbulent boundary layer, the gradient of  pressure in the radial direction is expected to be larger than in the vertical direction by a factor proportional to the spreading rate of the flow.
The DNS data confirms that this is the case (Figs \ref{fig:pressure}(d-f)). 


Figs \ref{fig:ss_second}(d-f) show the streamwise turbulent momentum and buoyancy flux. Whilst the vertical turbulent momentum flux is more or less identical for cases J, F and P, the buoyancy profile differs significantly between the three subplots.
Clearly, an increase in the value of $\Gamma$ increases the vertical turbulent buoyancy flux, as well as the radial buoyancy flux (Figs \ref{fig:ss_first}(d-f)).
A similar trend is observable in the turbulence buoyancy variance (Figs \ref{fig:ss_second}(g-i)).
Note that given a sufficient vertical extent of the domain, we expect both $\overline{w'b'}$ and $\overline{b'b'}$ for simulation F to increase to levels observed in simulation P.

To provide further evidence of the similarity of the turbulence statistics in plumes and jets it is instructive to calculate the invariants of the anisotropy tensor \citep{Pope2000}
\begin{equation}
  \label{eq:bij}
  b_{ij} = \frac{\overline{u_i'u_j'}}{2e} - \frac{1}{3} \delta_{ij},
\end{equation}
where $e=\frac{1}{2} \overline{u_i'u_i'}$ is the turbulence kinetic energy and $\delta_{ij}$ is the Kronecker delta.
As the turbulence is incompressible, one invariant of $\mathbf{b}$ is zero, and the other two, denoted $\xi$ and $\eta$, are defined via
$\textnormal{Tr}(\mathbf{b}^2) \equiv 6 \xi^2$ and $\textnormal{Tr}(\mathbf{b}^3) \equiv 6 \eta^3$, where Tr denotes the tensor trace.
The invariants of $\mathbf{b}$ cannot take any value; realisable flows are confined to a region of the $\xi-\eta$ space commonly known as the Lumley triangle\cite{Pope2000}.

\begin{figure}
\centering
\includegraphics{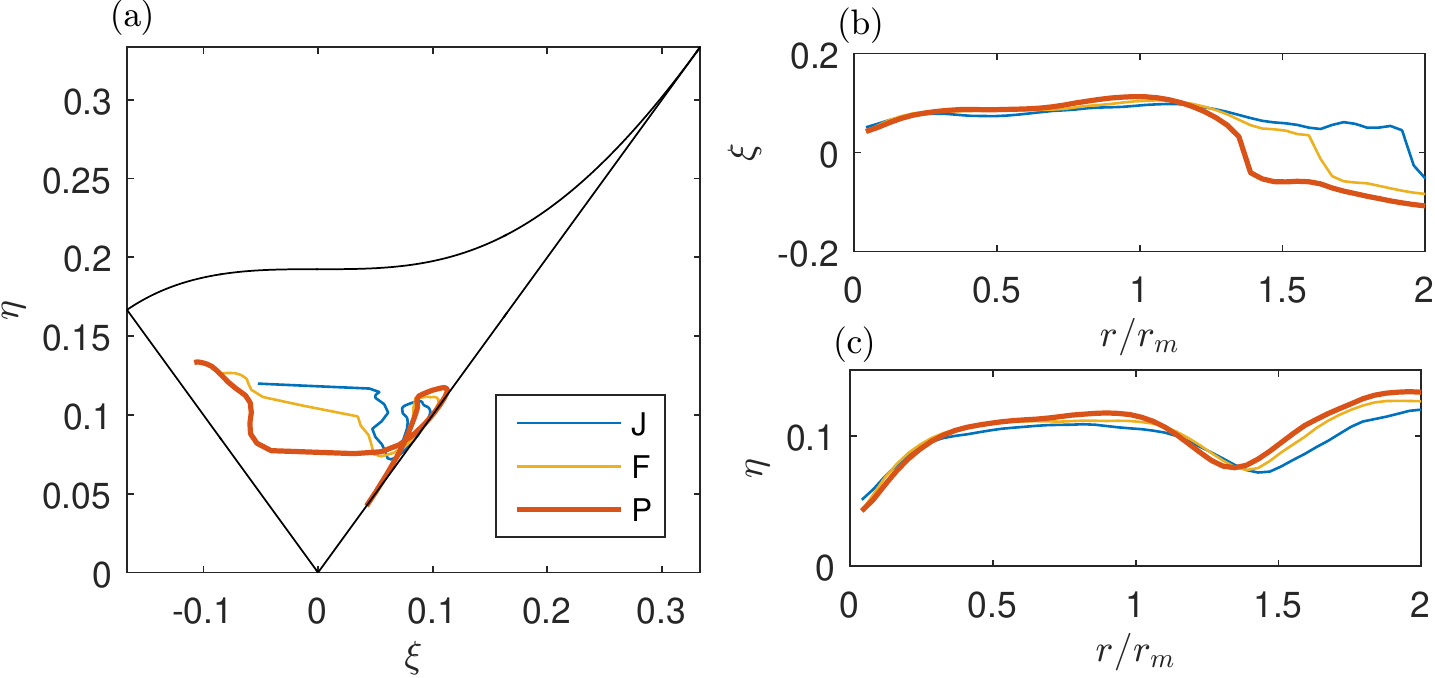}
\caption{Invariants of the anisotropy tensor \eqref{eq:bij} for the jet, forced plume and plume. (a) Plot in $\xi-\eta$ space, together with the Lumley triangle. (b) Dependence of $\xi$ on $r/r_m$. (c) Dependence of $\eta$ on $r/r_m$.}
\label{fig:lumley}
\end{figure}

The invariants are calculated as follows. The second-order statistics shown in
Figs \ref{fig:ss_first} and \ref{fig:ss_second} are averaged over the range
$20<z/r_m<50$, after which $\xi$ and $\eta$ are calculated as a function of
$r/r_m$.  Figs \ref{fig:lumley}(b, c) show, respectively, the profiles of
invariants $\eta$ and $\xi$ as a function of $r/r_m$.  It is evident that the profiles for J, F and P are nearly indistinguishable for $r/r_m<1.5$, providing further
evidence that turbulence in jets and plumes is similar.  In the $\xi-\eta$
plane (Fig. \ref{fig:lumley}(a)), the data is close to the $\xi=\eta$ line,
which is indicative of axisymmetric turbulence with one large eigenvalue,
i.e.\ rod-like turbulence.  Interestingly, at the edge of the jet/plume, $\xi$
changes very rapidly from positive to negative.  For plumes, the crossover
appears to happen closer to the centreline than for the jet.  Thus, near the
plume edge, the average picture of the turbulence resembles axisymmetric
turbulence with one small eigenvalue, i.e. disk-like turbulence.  These
observations are in agreement with the laboratory experiments of
\citet{Hussein1994}, which were presented in terms of the $(\xi,\eta)$
invariants in \citet{Kuznik2011}.

Consideration of the vertical gradient $\partial\overline{w}/\partial z$
provides a possible explanation for why the point at which turbulence changes
from being dominated by one component (the core region) to two components (the
edge of the flow) differs in jets compared with plumes. Noting that
$w_{m}\sim z^{-1}$ in jets, whereas $w_{m}\sim z^{-1/3}$ in plumes, the point
at which $\partial\overline{w}/\partial z=0$ occurs at larger values of
$r/r_{m}$ in jets than it does in plumes. Likening the flow with a diverging
(core region, $\partial\overline{w}/\partial z<0$) or converging (edge region,
$\partial\overline{w}/\partial z>0$) nozzle, one would therefore expect the
point of transition between one-component and two-component regimes,
respectively, to be affected by differences in the point at which $\partial\overline{w}/\partial z$ changes sign.

\subsection{Profile coefficients} 
\label{sec:profilecoefs}

Profile coefficients encapsulate integrated information about mean and turbulent fluxes of momentum, buoyancy, mean kinetic and turbulence production.
In classic integral descriptions of the plume equations \cite{Morton1956}, the profile coefficients are generally assumed to be either unity or zero.
However, preserving information about profile shapes is crucial in the description of unsteady jets and plumes \cite{Craske2015a, Craske2016, Woodhouse2016}, and is also the key to decomposing entrainment into its various processes.
The profile coefficients for momentum ($\beta$), buoyancy ($\theta$), energy ($\gamma$) and turbulence production ($\delta$) are given by, respectively:

\begin{equation}
\begin{aligned}
\beta_m & \equiv \frac{M}{w_m^2 r_m^2} \equiv 1, &
\beta_f, & \equiv \frac{2}{w_m^2 r_m^2} \int_0^\infty \AV{w'^2} r \d r, &
\beta_p & \equiv \frac{2}{w_m^2 r_m^2}  \int_0^\infty \AV{p} r \d r, \\
\gamma_m &\equiv \frac{2}{w_m^3 r_m^2} \int_0^\infty \AV{w}^3 r \d r, &
\gamma_f &\equiv \frac{4}{w_m^3 r_m^2} \int_0^\infty \AV{w} \AV{w'^2} r \d r, &
\gamma_p &\equiv \frac{4}{w_m^3 r_m^2} \int_0^\infty \AV{w} \AV{p}  r \d r, \\
\delta_m &\equiv \frac{4}{w_m^3 r_m} \int_0^\infty \AV{w'u'} \frac{\partial \AV{w}}{\partial r} r \d r, &
\delta_f &\equiv \frac{4}{w_m^3 r_m} \int_0^\infty \AV{w'^2} \frac{\partial\AV{w}}{\partial z} r \d r, &
\delta_p &\equiv \frac{4}{w_m^3 r_m} \int_0^\infty \AV{p} \frac{\partial \AV{w}}{\partial z} r \d r, \\
\theta_m &\equiv \frac{F}{w_m b_m r_m^2}, &
\theta_f &\equiv \frac{2}{w_m b_m r_m^2} \int_0^\infty \AV{w'b'} r \d r. &
\end{aligned}
\label{eq:profilecoefs}
\end{equation}
The total momentum flux is given by $\beta_g M$, where $\beta_g = \beta_m+\beta_f+\beta_p$. Similarly, $\theta_g$ is associated with the total buoyancy flux, $\gamma_g$ with the total energy flux and $\delta_g$ with the total turbulence production (including pressure redistribution).
Profile coefficients $\beta$ and $\theta$ show up naturally upon radial integration of the Reynolds-averaged volume, vertical momentum and buoyancy equations of a high Reynolds number flow in a neutral environment \citep{vanReeuwijk2015}
\refstepcounter{equation}\label{eq:plumeeqs}
\begin{equation}
\tag{{\theequation}a-c}
\frac{1}{Q} \OD{Q}{\zeta} = 2 \alpha, \quad\quad
\frac{1}{M}\OD{}{\zeta} \left(\beta_{g}M \right) = \Ri, \quad\quad
\frac{1}{F}\OD{}{\zeta} \left(\frac{\theta_{g}}{\theta_m} F \right)= 0.
\end{equation}
These equations reduce to the classic plume equations \cite{Morton1956} on setting $\beta_g = 1$ and $\theta_g = \theta_m = 1$. 
Furthermore, we note that $\Ri = 0$ by definition for the jet, implying that the evolution of $F$ and $M$ are uncoupled (and that $F$ in that case corresponds to a passive scalar flux).
Similarly, $\gamma$ and $\delta$ emerge naturally from integration of the mean kinetic energy equation:
\begin{equation}
\frac{Q}{M^{2}} \OD{}{\zeta} \left( \gamma_{g}\frac{M^{2}}{Q} \right) = \delta_{g} +2 \theta_m \Ri.
\label{eq:E1}
\end{equation}

\begin{figure}
\centering
\includegraphics{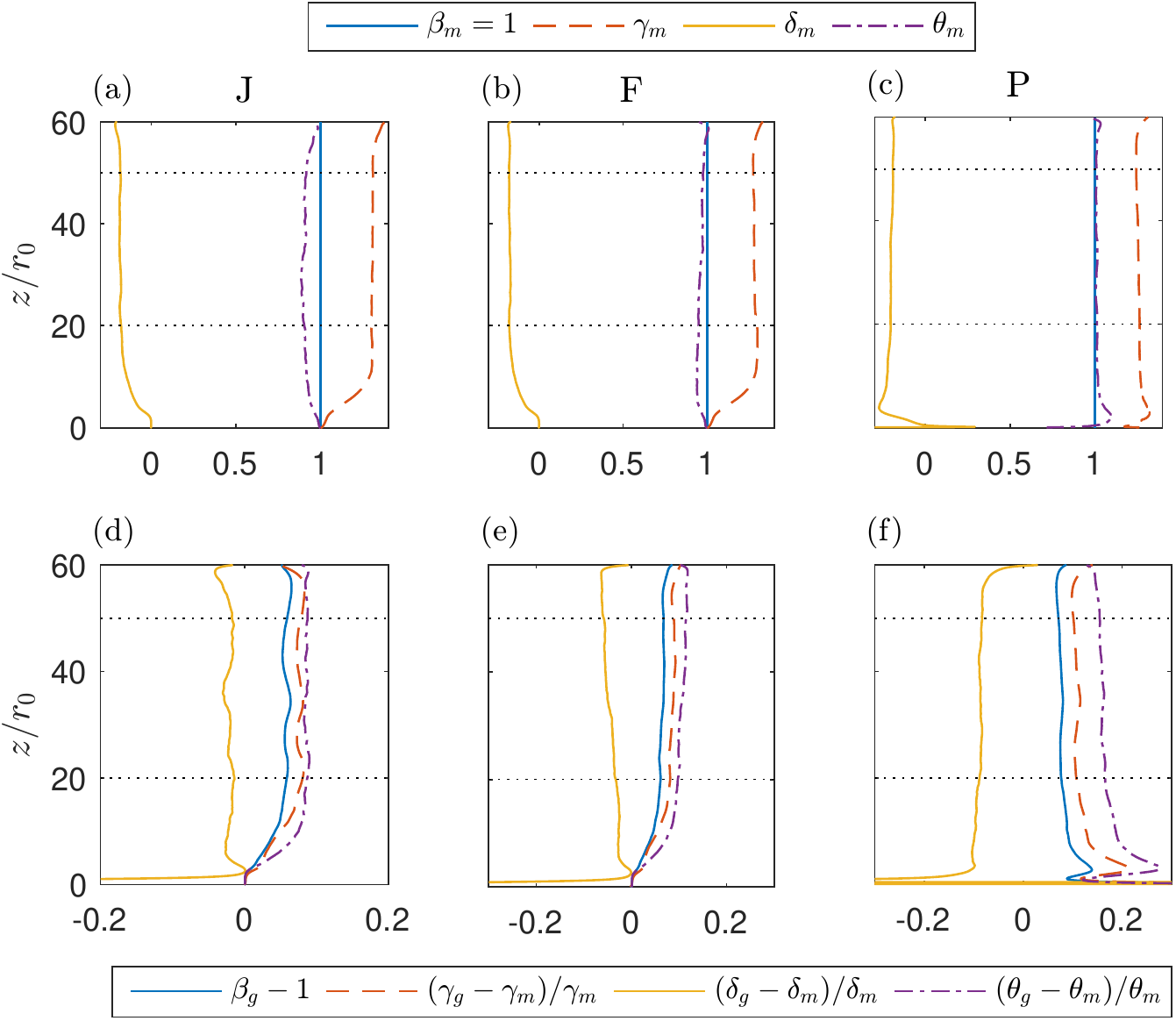}
\caption{(a-c) Mean profile coefficients. (d-f) The relative contribution of turbulence and pressure to the dimensionless coefficients. (a,d) Simulation J. (b,e) Simulation F. (c,f) Simulation P. The dashed lines indicate the averaging interval $20<z/r_0<50$ used for the profile coefficients displayed in Table \ref{tab:profcoefs}.}
\label{fig:profilecoefs}
\end{figure}

Fig.\ \ref{fig:profilecoefs} shows the profile coefficients as a function of $z$. The coefficients associated with the mean flow, $\beta_m$, $\gamma_m$, $\delta_m$ and $\theta_m$, are shown in Figs \ref{fig:profilecoefs}(a-c). 
There are large variations in the profile coefficients in the near field, which are due to changes in the velocity and buoyancy profiles as the jet/plume develops; indeed, the largest changes occur over a small region $z/r_0<5$, for the plume even closer to the source ($z/r_0<3$).
However, for larger $z/r_0$ the coefficients become constant, which is consistent with self-similarity.

The average values of the profile coefficients over the interval $20<z/r_0<50$ are presented in Table \ref{tab:profcoefs}.
The dimensionless buoyancy flux $\theta_m$ is less than unity for the jet, implying that the spread of the buoyancy field exceeds the spread of the velocity field.
This can be shown by assuming a Gaussian form for the velocity and buoyancy profiles
\begin{equation}
  \AV{w}= 2 w_m \exp\left(-2 \frac{r^2}{r_m^2}\right), \quad\quad
  \AV{b}= 2\frac{b_m}{\varphi^2} \exp\left(-2 \frac{r^2}{\varphi^2 r_m^2}\right),
  \label{eq:wbGaussian}
\end{equation}
where $\varphi r_m$ is the characteristic width of the buoyancy profile and $\varphi$ is the ratio of the buoyancy to velocity radii.
These profiles are consistent with the definitions $\beta_m = 1$ and $B=b_m r_m^2$, and evaluation of the profile coefficient for the mean energy flux results in $\gamma_m = 4/3$. The buoyancy flux is given by $
  F = 2 \int_0^\infty \AV{w} \AV{b} r \d r = \frac{2}{\varphi^2+1} w_m b_m r_m^2$.
By substituting this expression into the definition of profile coefficient $\theta_m$ \eqref{eq:profilecoefs}, it directly follows that 
\begin{equation}
\theta_m = \frac{2}{\varphi^2+1}.
\end{equation}
For the plume, $\theta_m \approx 1$, implying that $\varphi\approx 1$ also. 
The value of $\theta_m$ for the forced plume tends to become closer to unity with increasing $z$.
The dimensionless turbulence production $\delta_m$ shows differences of the order of 10\% between the jet and the plume (see also Table \ref{tab:profcoefs}), which is too small to explain the observed differences in $\alpha$ (see section \ref{sec:alpha}).

 \setlength{\tabcolsep}{0.3cm}
\begin{table}
  \begin{center}
\def~{\hphantom{0}}
  \begin{tabular}{l|ccc}
& J
& F
& P \\
\hline
$\beta_f$&0.151&0.149&0.183\\
$\beta_u$&0.095&0.088&0.106\\
$\beta_v$&0.102&0.095&0.110\\
$\beta_p$&-0.093&-0.084&-0.107\\
$\beta_g$&1.058&1.065&1.076\\
$\gamma_m$&1.306&1.282&1.256\\
$\gamma_f$&0.276&0.267&0.319\\
$\gamma_p$&-0.175&-0.156&-0.183\\
$\gamma_g$&1.406&1.393&1.391\\
$\delta_m$&-0.184&-0.175&-0.201\\
$\delta_f$&0.006&0.016&0.038\\
$\delta_p$&-0.002&-0.008&-0.021\\
$\delta_g$&-0.180&-0.167&-0.184\\
$\theta_m$&0.901&0.964&1.011\\
$\theta_f$&0.078&0.103&0.162\\
$\theta_g$&0.979&1.067&1.172
 \end{tabular}
  \end{center}
  \caption{Average profile coefficients over the interval $20<z/r_0<50$.}
  \label{tab:profcoefs}
\end{table}

Figs \ref{fig:profilecoefs}(d-f) show the relative contribution of turbulence and pressure terms to the total, which are neglected in classic plume theory.
Gradual changes can be observed in the far-field which are caused by the fact that the second-order statistics require a greater vertical distance to become fully self-similar than the first-order statistics. 
Indeed, \citet{Wang2002} observed that full self-similarity of the turbulence statistics did not occur before $z/r_0\approx 100$, which is nearly twice the vertical extent of our domain. 
However, it is clear that in general, the influence of turbulence and pressure is less than 10\% of the mean value, which partially explains why plume theory provides such robust predictions for plume behaviour.
The largest deviations between mean and total are found in $\theta$, the dimensionless buoyancy flux, which for plumes is as high as 20\%, consistent with literature \cite{Linden2003, Devenish2010}.
Here, we would like to point out that  $\theta_f$ is a source of systematic error in laboratory experiments where the (total) buoyancy flux is usually determined \emph{a priori} (nozzle volume flux $\times$ buoyancy).
However, plume theory only considers means, and the mean buoyancy flux is about 20\% less than the total buoyancy flux.
Indeed, we find good agreement of the DNS data with the classic solutions of plume theory only by explicitly calculating the mean buoyancy flux.

\subsection{Decomposing the entrainment coefficient}
\label{sec:alpha}
\begin{figure}
\centering
\includegraphics{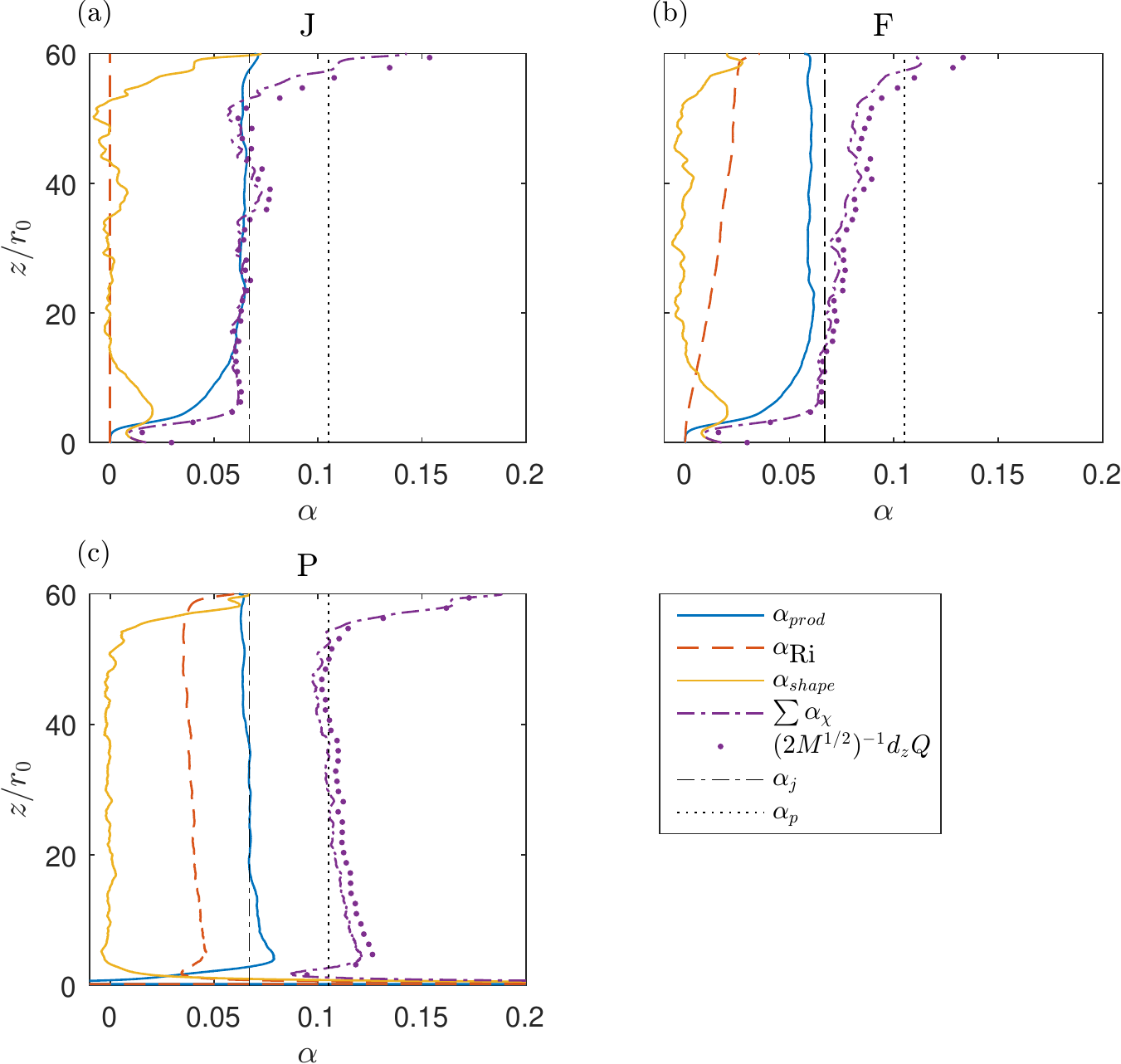}
\caption{Evolution of the contribution to entrainment due to turbulent kinetic energy production $\alpha_{prod}$, buoyancy $\alpha_{\Ri}$ and departure from self-similarity $\alpha_{shape}$, as a function of $z$. Note that in the legend, $\sum \alpha_\chi = \alpha_{prod} + \alpha_{\Ri} + \alpha_{shape}$. }
\label{fig:alpha_decomp}
\end{figure}

As shown in \citet{vanReeuwijk2015}, taking \eqref{eq:alphadef} as a definition of $\alpha$, and using \eqref{eq:E1} and (\ref{eq:plumeeqs}b), $\alpha$ can be decomposed as:

\begin{equation}
\begin{split}
\alpha = \underbrace{-\frac{\delta_g}{2\gamma_g}}_{\alpha_{prod}}
+\underbrace{\left(\frac{1}{\beta_g}-\frac{\theta_m}{\gamma_g}\right)\Ri}_{\alpha_{\Ri}}
+\underbrace{\OD{}{\zeta} \left( \log \frac{\gamma_g^{1/2}}{\beta_g} \right)}_{\alpha_{shape}}.
\end{split}
\label{eq:alpha2}
\end{equation}
The entrainment relation (\ref{eq:alpha2}) quantifies the contribution to $\alpha$ of turbulence production $\alpha_{prod}$, mean buoyancy $\alpha_{\Ri}$ and changes in profile shape $\alpha_{shape}$.
The vertical evolution of the individual contributions to $\alpha$, as well as the direct estimate of $\alpha$ using \eqref{eq:alphadef} and the estimate of $\alpha$ using $r_m$ (Table \ref{tab:sim}) are plotted in Fig.\ \ref{fig:alpha_decomp}.
The three estimates of $\alpha$ are in good agreement with each other, demonstrating the consistency of the data with the underlying integral equations.
The analysis of data from the plume literature carried out in \citet{vanReeuwijk2015} (VRC15) highlighted that $\delta_m$, and thus $\alpha_{prod}$, was approximately identical in jets and plumes.
This is convincingly confirmed in Fig.\ \ref{fig:alpha_decomp}(c), as $\alpha_{prod}$ matches closely with the value of $\alpha_j$ inferred from the jet data.
For the forced plume, $\alpha_{prod}$ is slightly lower than $\alpha_j$ but remains in good agreement.
The mean-flow contribution of buoyancy to $\alpha$ is constant for simulation P, and has a magnitude of $2 \alpha_j/3$.
For simulation F, $\alpha_{\Ri}$ can be observed to increase with height.

The term $\alpha_{shape}$ will only be non-zero when the profiles of first- and second-order statistics change in shape, i.e. when the profiles are not self-similar.
Non-self-similar behaviour is dominant in the near field, where the flow transitions to turbulence and the mean profiles attain their Gaussian shapes. 
The near-field region, within which $\alpha_{shape}$ is different from zero, extends up to about 15 source diameters for the jet and the forced jet, and only for about 5 source diameters for the plume.

\begin{figure}
\includegraphics{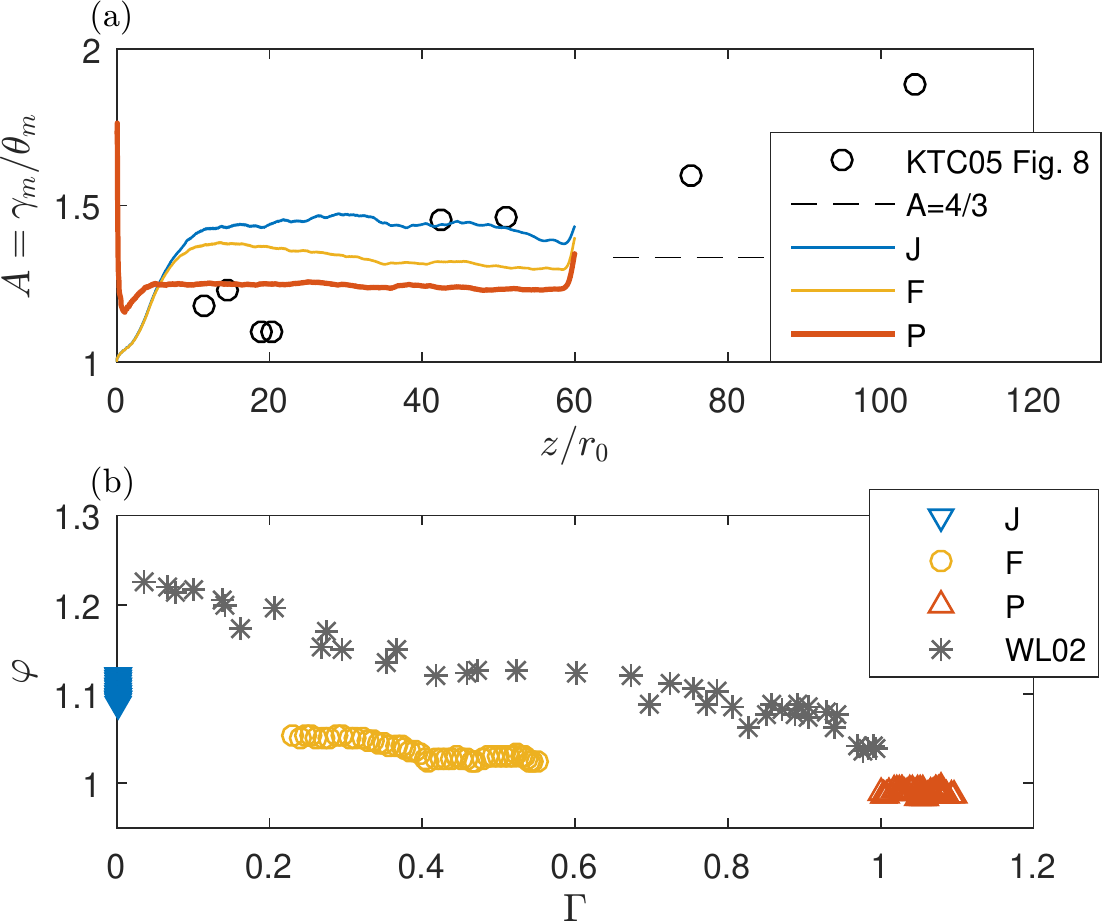}
\centering
\caption{Exploration of similarity drift. (a) $A=\gamma_m / \theta_m$ as a function of $z/r_0$. (b) $\varphi$ as a function of $\Gamma$.}
\label{fig:similaritydrift}
\end{figure}

Next, we explore the concept of similarity drift, which pertains to a possible variation in $z$ of the ratio of buoyancy to velocity profile width $\varphi(z)$.
The concept of similarity drift can be traced back to \citet{Kaminski2005} (KTC05), who derived an entrainment relation that contains a term of the form
\begin{equation}
\label{eq:alphaKaminski}
\alpha_e = \hdots + \frac{1}{2} R \frac{\d}{\d z} \log A,
\end{equation}
where $R$ is a typical radius,
$A=\gamma_m / \theta_m = \gamma_m (1+ \varphi^2) / 2$ and $\alpha_e$ is an
entrainment coefficient that is related\citep{vanReeuwijk2015}, but not identical to
$\alpha$ ($\alpha_e$ uses non-standard characteristic scales in KTC05,
implying that the $\alpha_{shape}$ in the entrainment relation in terms of
$\alpha$ \eqref{eq:alpha2} is independent of $\theta$). Hence,
\eqref{eq:alphaKaminski} indicates that changes in $A$, e.g.\ because of a
drift $\varphi = \varphi(z)$ will have a non-zero contribution to $\alpha_e$.
In KTC05, the value $A$ was calculated for published data which, despite significant scatter, showed an increasing trend of $A$ with the distance from the source.

Fig.\ \ref{fig:similaritydrift}(a) shows the experimental data collected from Fig.\ 8 in KTC05 together with the new DNS data set discussed in this
article. Unlike the experimental data, the DNS data does not imply that $A$ varies as a function of $z$. 
Indeed, it is unclear what physical mechanism could be responsible for producing similarity drift. 
Full self-similarity of the process results from an asymptotically small dependence on the source
conditions and ambient conditions that scale in the same way as the local
behaviour of the plume. We therefore suggest that the similarity drift
observed in experiments is caused by the absence of an ideal undisturbed,
unbounded ambient environment (including confinement effects) or a persistent
dependence of the process on source conditions.

The DNS and WL02 data suggest a relation between $\varphi$ and $\Gamma$, see
Fig.\ \ref{fig:similaritydrift}(b).  As for Fig.\ \ref{fig:alphaRi}, the DNS and WL02 data show that $\varphi$ is a decreasing function of $\Gamma$,
tending to $\varphi\approx 1$ at $\Gamma=1$.  The $\Gamma$-dependence is more
pronounced for the WL02 data than the DNS data, the reason for which is unclear.

\subsection{Turbulent transport}
\label{sec:nuT}

\begin{figure}
\centering
\includegraphics{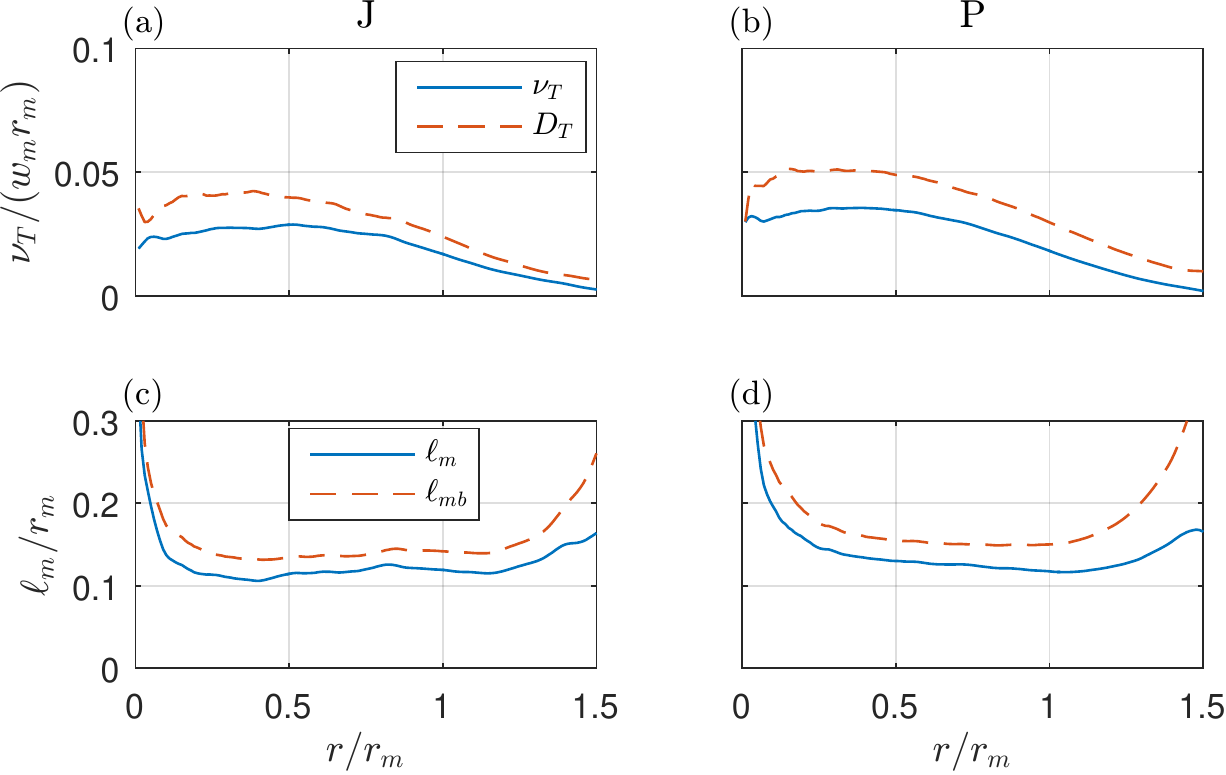}
\caption{Radial profiles of $\nu_T$ and $D_T$: (a) for the jet and (b) and for the pure plume. Radial mixing length radial profiles: (c) for the jet and (d) for the pure plume.}
\label{fig:nuTDT}
\end{figure}

The turbulent radial transport of streamwise momentum $\overline{u'w'}$ and buoyancy $\overline{u'b'}$ are crucial in determining the profile shape and entrainment behaviour of jets and plumes.
These quantities can be related to the mean fields using the gradient diffusion hypothesis, i.e.
\begin{equation}
 \label{eq:GDH}
 \overline{u'w'} = -\nu_T \frac{\partial \overline{w}}{\partial r}, \quad \quad 
 \overline{u'b'} = -D_T \frac{\partial \overline{b}}{\partial r}.
\end{equation}
%
These quantities were computed using $\nu_T / (w_m r_m) = - f_{uw} / f_w'$ and $D_T / (w_m r_m) = - f_{ub}/f_b'$, where the similarity functions $f_\chi$ are the averages of those presented in Fig.\ \ref{fig:ss_first} and the prime denotes differentiation with respect to $\eta$.
The results are shown in Figs \ref{fig:nuTDT}(a,b) for the jet and plume, respectively.
The radial distributions of $\nu_T$ and $D_T$ have a similar shape, with $D_T$ systematically higher than $\nu_T$ for both the jet and the plume.
The values for $\nu_T$ and $D_T$ are slightly higher for the plume than for the jet.

The profiles for $\nu_T$ and $D_T$ show substantial variations over the interval $0<r/r_m<1$.
A Prandtl mixing length model \cite{Schlichting1960} with mixing lengths for momentum and buoyancy of the form $\ell_m = a_w r_m$ and $\ell_{mb} = a_b r_m$, resulting in 
\begin{equation}
  \label{eq:lm}
  \frac{\nu_T}{w_m r_m} = a_w^2 |f'_w|, \quad \quad
  \frac{D_T}{w_m r_m} = a_b^2 |f'_w|,
\end{equation}
provides values of $\ell_m / r_m \equiv a_w$ and $\ell_{mb} / r_m \equiv a_b$, that are roughly constant in the core region (Figs \ref{fig:nuTDT}(c,d)).
Very close to the centreline, the mixing length becomes very large because $|f'_w|$ and $|f'_b|$ tend to zero.
For $r/r_m>1$, the mixing length concept does not work well, which we attribute to intermittency effects associated with the plume edge.
The typical values for $a_w$ and $a_b$ over the region $0.3 < r/r_m < 1.0$ are presented in Table \ref{tab:sim}.
Estimates of the mixing length show a remarkable agreement with the experimental results recently presented by \citet{Ezzamel2015}, who estimated the Eulerian integral length scale of the two-point velocity statistics (their figure 15). 
In particular, note that the measurements revealed almost constant values of the Eulerian integral length in the core of the plume, for both jets and plumes. 

\begin{figure}
\centering
\includegraphics{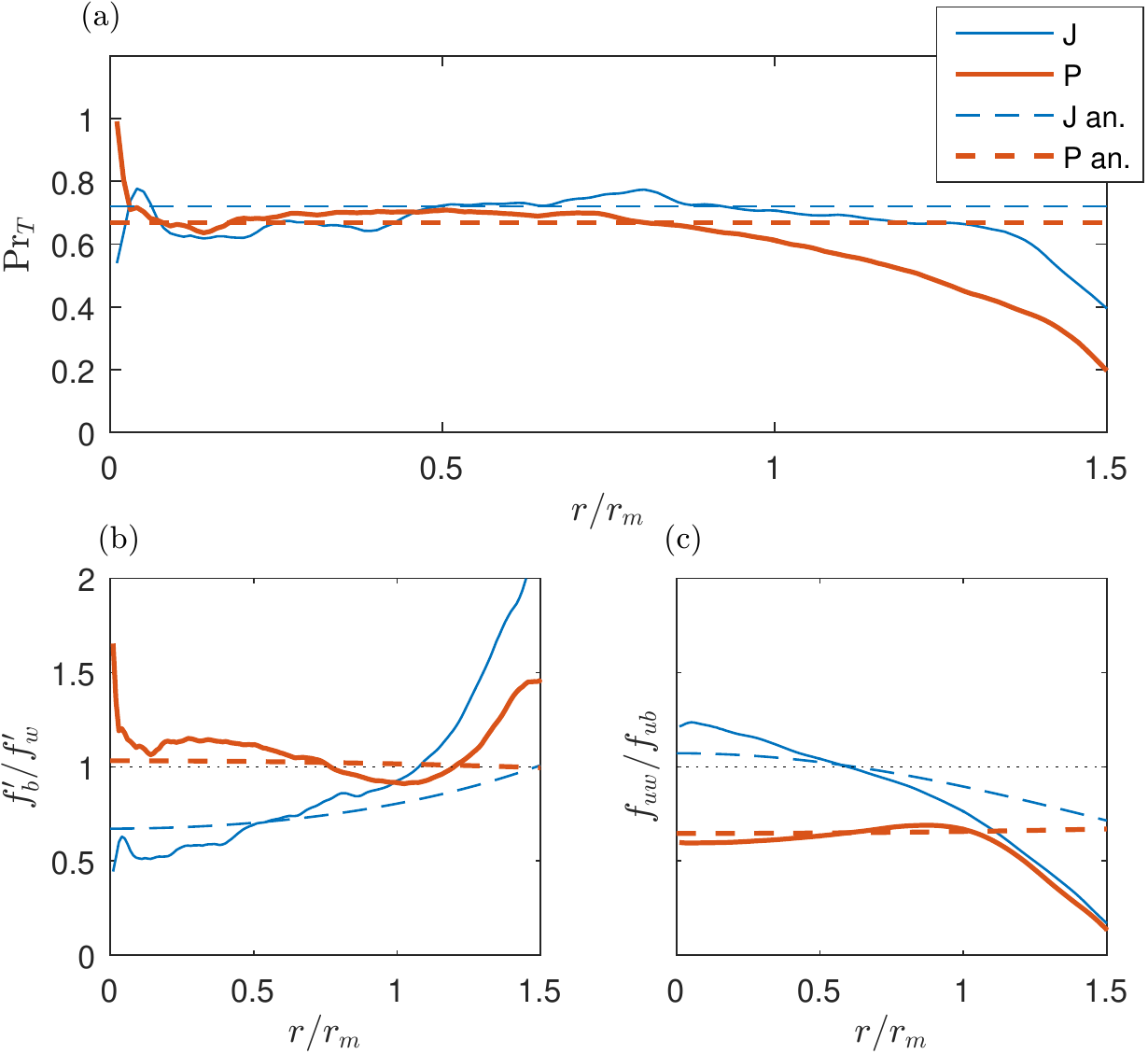}
\caption{ Radial profiles (a) of the turbulent Prandtl number $\Pr_T $, (b) of the ratio of the similarity functions $f_b'/f_w'$ and (c) $f_{uw}/f_{ub}$ (see text). Solid lines: DNS data. Dashed lines: analytical predictions \eqref{eq:ratios}.
}
\label{fig:PrT}
\end{figure}

The turbulent Prandtl number $\Pr_T$ is a quantity of great relevance because of its extensive use in turbulence modelling.
By substituting \eqref{eq:GDH} into \eqref{eq:PrT}, one obtains
\begin{equation}
\label{eq:PrT2}
\Pr_T= \frac{\nu_T}{D_T} = \frac{f_{uw}}{f_{ub}} \frac{f'_b}{f'_w}.
\end{equation}
Thus, $\Pr_T$ can be thought of as the product of two ratios: 1) the ratio of the radial turbulent fluxes $f_{uw}/f_{ub}$ and 2) the ratio of gradients of the mean buoyancy and velocity $f_b'/f_w'$.
The turbulent Prandtl number, plotted in Fig. \ref{fig:PrT}, is almost constant over the entire cross section with values in the range $0.6$ - $0.8$.
The average value $\langle \Pr_T \rangle$ over the interval $0.3<r/r_m<1.0$ is 0.72 for the jet simulation and 0.67 for the plume simulation (see also Table \ref{tab:sim}).
Thus, the estimates of $\langle \Pr_T \rangle$ are remarkably close, despite the effect of buoyancy on the plume's behaviour.
Shown in Fig.\ \ref{fig:PrT}(b) is the ratio $f_b'/f_w'$.
For the plume, the ratio is approximately unity, but for the jet it is significantly lower due to the fact that $\theta_m<1$ and thus $\varphi>1$.
The ratio $f_{uw}/f_{ub}$, shown in Fig.\ \ref{fig:PrT}(c), is approximately constant for the plume with a value of about 0.6. For the jet, $f_{uw}/f_{ub}$ decreases slowly with an average value of about 1.  

Thus, although $\Pr_T$ is very similar for plumes and jets, the reason is different: for jets it is caused primarily by $f_b'/f_w'$ which is associated with the ratio of widths $\varphi$, and for the plume primarily by the turbulent flux ratio $f_{uw}/f_{ub}$, see also Fig.\ \ref{fig:ss_first}.
This can be made explicit by evaluating the ratios (by substituting the Gaussian profiles for $f_w=\overline{w}/w_m$ and $f_b=\overline{b}/b_m$ \eqref{eq:wbGaussian}) into \eqref{eq:GDH}, \eqref{eq:lm}, resulting in

\begin{equation}
  \label{eq:ratios}
  \frac{f_b'}{f_w'} = \frac{1}{\varphi^4} \exp\left( -\frac{\varphi^2-1}{\varphi^2} \frac{r^2}{r_m^2} \right),
  \quad\quad
  \frac{f_{uw}}{f_{ub}} = \varphi^4 \frac{a_w^2}{a_b^2} \exp\left( \frac{\varphi^2-1}{\varphi^2} \frac{r^2}{r_m^2} \right),
\end{equation}
noting that $(\varphi^2-1)/\varphi^2 = (2 - 2 \theta_m) / (2 - \theta_m)$.
The product of these two terms evaluates to $\langle \Pr_T \rangle = a_w^2/a_b^2$, consistent with \eqref{eq:lm}.
Eq.\ \ref{eq:ratios} shows that the amplitude of the ratio $f_b'/f_w'$ is solely determined by the value of $\varphi$.
The amplitude of the ratio $f_{uw}/f_{ub}$ is determined both by $\varphi$ and the ratio of mixing lengths $a_w/a_b$.
The theoretical predictions of \eqref{eq:ratios}, using parameter values for $a_w$, $a_b$ from Table \ref{tab:sim} and $\theta_m$ from Table \ref{tab:profcoefs} are plotted in Fig.\ \ref{fig:PrT} with dashed lines.
The results agree quite well in the interval $0<r/r_m<1$, both in terms of the amplitude and in the trend.
Near the plume edge, it is clear that the mixing lengths and Gaussians do not describe the behaviour.

Previous authors \cite{Ezzamel2015} have suggested that a spatially averaged
(over the radial plume section) turbulent Prandtl number
$\langle \Pr_T \rangle$ can be inferred from the ratio of the plume radii
$r_m$ and $r_b$, estimated through a Gaussian fit of the radial profiles of
mean vertical velocity and buoyancy, respectively. For jets this approach is
valid because, to leading order, the scalar field and the vertical velocity
field essentially obey the same similarity equations, which state that radial
mixing must balance the divergence in the vertical flux. As noted
previously\cite{Chen1976}, the ratio of $r_m$ and $r_b$ can be obtained via
the substitution of Gaussian profiles into the similarity
equations. Evaluation of the resulting balance on the centreline of the flow
allows one to relate $D_{T}$ to $r_{b}$ and $\nu_{T}$ to
$r_{m}$. Equivalently, one can view the problem in a moving frame of
reference, in which $z^{2}\propto t$, and apply the classic relation for
diffusion, which predicts that $r_{b} \propto \sqrt{t D_T}$ and
$r_{m} \propto \sqrt{t \nu_T}$. Both approaches result in
$\langle \Pr_T \rangle = \varphi^{-2}$. For jets, we observe that
$\varphi\approx 1.1$ and therefore would expect
$\langle \Pr_T \rangle \approx 0.8$, which is reasonably consistent with
Fig. \ref{fig:PrT}(a). In the case of plumes, however, the analysis described
above is not appropriate, unless one accounts for the additional term arising
from buoyancy in the governing momentum equation.  Indeed, our results
indicate values of $\langle \Pr_T \rangle$ which are systematically lower than
unity in plumes (see e.g.\ Fig. \ref{fig:PrT}(a)), in spite of the fact that
$\varphi \approx 1$. 

\section{Conclusions}
\label{sec:conclusions}

The dynamics and transport properties of a turbulent pure jet, a pure plume and a forced plume were examined using high-fidelity direct numerical simulations. The motivation for this work, the numerical analogue of the experimental study by \citet{Ezzamel2015}, was specifically to shed light on the physical processes linking turbulent transport and entrainment.

The detailed spatial resolution of the DNS allowed the effectiveness of turbulent transport to be quantified, e.g.\ via turbulent diffusion coefficients and the dilution of fluid in the plume/jet with the ambient. For the forced plume, within which the flow dynamically adjusts towards a pure-plume behaviour asymptotically with height, of particular relevance was the vertical variation of the entrainment coefficient $\alpha$, numerous models having been proposed to capture this variation. Our results support the \citet{Priestley1955} entrainment model \eqref{eq:alphaPB} and show that, beyond a near-source region (specifically for $z/r_0 \gtrsim 20$), the entrainment coefficient is a function only of the local Richardson number.

By decomposing $\alpha$  (see \eqref{eq:alpha2}) into contributions due to turbulence production, to buoyancy and to shape effects, we show that the production of turbulence due to shear (as represented by the dimensionless quantity $\delta_m$) is practically identical for jets and for plumes, which is indeed the assumption underlying \eqref{eq:alphaPB}. Moreover,  since the \textit{turbulent} component of entrainment has been shown to be unaltered by buoyancy \citep{vanReeuwijk2015}, this confirms that $\alpha$ is larger for plumes than for jets due to entrainment associated with \textit{mean} flow processes.

The fact that the production of turbulence due to shear takes approximately the same value for jets and plumes suggests that their turbulence structure is quite similar, despite the absence of buoyancy in a jet.  The second-order statistics $\overline{u'u'}$, $\overline{v'v'}$ and $\overline{w'w'}$ indeed suggest the turbulence levels are very similar.  The invariance of the turbulence anisotropy tensor confirms that turbulence in the core region of a jet/plume is practically indistinguishable.  There is, however, evidence of clear distinctions between the structure of a jet and a plume. For example, whilst there is a transition from rod-like to disk-like turbulence moving radially outward from the centreline, this transition occurs closer to the centreline in a plume; these distinctions are believed to be linked with vertical velocity gradients $\partial \overline{w} / \partial z$. Further differences between jets and plumes exist in the second-order scalar statistics, such as $\overline{w'b'}$ and $\overline{b'b'}$. Analysis of the budgets for these quantities would indicate how such differences can exist between flows whose dynamics are similar, and would therefore make a valuable contribution to an overall understanding of turbulence in free-shear flows.

In agreement with existing measurements, the turbulent Prandtl number is found to be almost identical for jets and plumes, taking a value of $\langle \Pr_T \rangle = 0.7$.
However, by writing this quantity as the ratio of turbulent fluxes and radial gradients of mean quantities, it becomes evident that for jets, the value of $\langle \Pr_T \rangle$ can be attributed to differences in the ratio of velocity to buoyancy profile widths $\varphi$, whereas for plumes, the value of $\langle \Pr_T \rangle$ is associated with the ratio of the turbulent radial transport of buoyancy and streamwise momentum.

The DNS data does not support the notion of similarity drift, and we conjecture that the observed variations in profile widths between experiments are possibly a result of confinement or other deviations from ideal boundary conditions.

\section*{Acknowledgments}
We acknowledge the UK Turbulence Consortium (grant number EP/L000261/1) and an
EPSRC ARCHER Leadership Grant for providing the computational resources
required to carry out the computations.  In addition, JC gratefully
acknowledges funding from the Engineering and Physical Sciences Research
Council (EPSRC) Doctoral Prize under grant number EP/M507878/1. 
PS gratefully acknowledges the support of Cambridge University's Engineering Department and Peterhouse College in June 2015.
 Supporting data for this article is available on request: please contact either
m.vanreeuwijk@imperial.ac.uk or civilsfluids@imperial.ac.uk.

\appendix

\section{Ambient-flow correction of the ESH15 data.}
\label{app:ESH15}
The purpose of this appendix is two-fold: 1) to correct the data of
\citet{Ezzamel2015} (ESH15) for vertical variation in the ambient flow; and 2)
to present the experimental data in terms of the notation used in this paper.

A significant part of the work in ESH15 was associated with the analysis of
$\alpha(z)$.  The $z$-dependence of $\alpha$ was determined in two ways: 1)
via volume conservation \eqref{eq:alphadef}; and 2) via the entrainment
relation \eqref{eq:alpha2} considering mean contributions and self-similarity
only, assuming Gaussian profiles ($\gamma_m=4/3$):
\begin{equation}
\label{eq:alphaMS}
\alpha = -\frac{3}{8}\delta_m + \left(1 - \frac{3}{4}\theta_m \right) \Ri.
\end{equation}
In ESH15, this relation was presented in terms of the relative plume width
$\varphi=\sqrt{2/\theta_m -1}$, the effective eddy viscosity
$\langle\widehat{\nu_T}\rangle = - \delta_m / (8 \sqrt{2})$ and the flux
balance parameter $\Gamma = 5\Ri/(8\alpha_p)$ (note that $\beta_g = 1$ as only means are considered), i.e.\ as
\begin{equation}
\label{eq:alphaESH15}
\alpha_G =3 \langle\widehat{\nu_T}\rangle + (2 \varphi^2-1)\frac{2 \alpha_{pG} \theta_m}{5}\Gamma.
\end{equation}
Here, $\alpha_G = \alpha/\sqrt{2}$ is the Gaussian entrainment coefficient and $\alpha_{pG} = \alpha_p/\sqrt{2}$ the Gaussian entrainment coefficient for a pure plume.
The prefactor for $\langle\widehat{\nu_T}\rangle$ is a factor two larger than reported in ESH15.  
Furthermore, the factor $\theta_m$ in the buoyancy contribution was not present in ESH15; this is caused by the inclusion of $\beta_{g}$ and $\theta_{m}$ in the flux balance parameter $\Gamma$ \eqref{eq:Gamma}.  
Indeed, denoting the classic flux balance parameter \citep{Morton1959} by
$\Gamma^*=5F Q^2/(8 \alpha_p M^{5/2})$, we have
$\Gamma^* = \beta_g \theta_m \Gamma$.

As discussed in ESH15, the measurements revealed a small but significant flow
in the ambient, caused by i) the diffusion of heat from the warm-air plume
source along the horizontal rigid wooden base plate within which the plume
nozzle was mounted, giving rise to vertical convective motion; and ii) the
seeding of the ambient with a stage smoke generator.  Indeed, the background
mean motion, whose vertical velocity we denote $\Delta w$, was clearly
captured by Particle Image Velocimetry (PIV) fields when measuring velocities
away from the plume perimeter in the lower regions of the domain -- a region
where the plume width was significantly smaller than the lateral extent of the
PIV field. Measurements indicated $\Delta w \approx 0.15$ $\mathrm{m s^{-1}} $
close to the source (whose radius is denoted $r_0$) for the jet-like, the forced and the pure-plume experiments, referred to as J, F and P, respectively.  
However, at larger vertical
distances above the source, the size of the PIV field did not permit
measurement of the (now significantly wider) plume or the ambient far beyond
the plume perimeter.  In ESH15 it was therefore assumed that the background
motion was uniform throughout the domain; hence $\Delta w = 0.15$
$\mathrm{m s^{-1}} $ was subtracted from the mean vertical velocities before
fitting the radial profiles with a Gaussian curve of the form
\begin{equation}\label{eqn:gauss}
\frac{\overline{w}(r,z)}{w_g(z)} =  \exp \left( \frac{-r^2}{r_g^2(z)} \right),
\end{equation}
where $w_g=2w_{m}$ denotes the plume centreline velocity and $r_g=r_{m}/\sqrt{2}$ the Gaussian plume radius.

\begin{figure}
\centering
\includegraphics{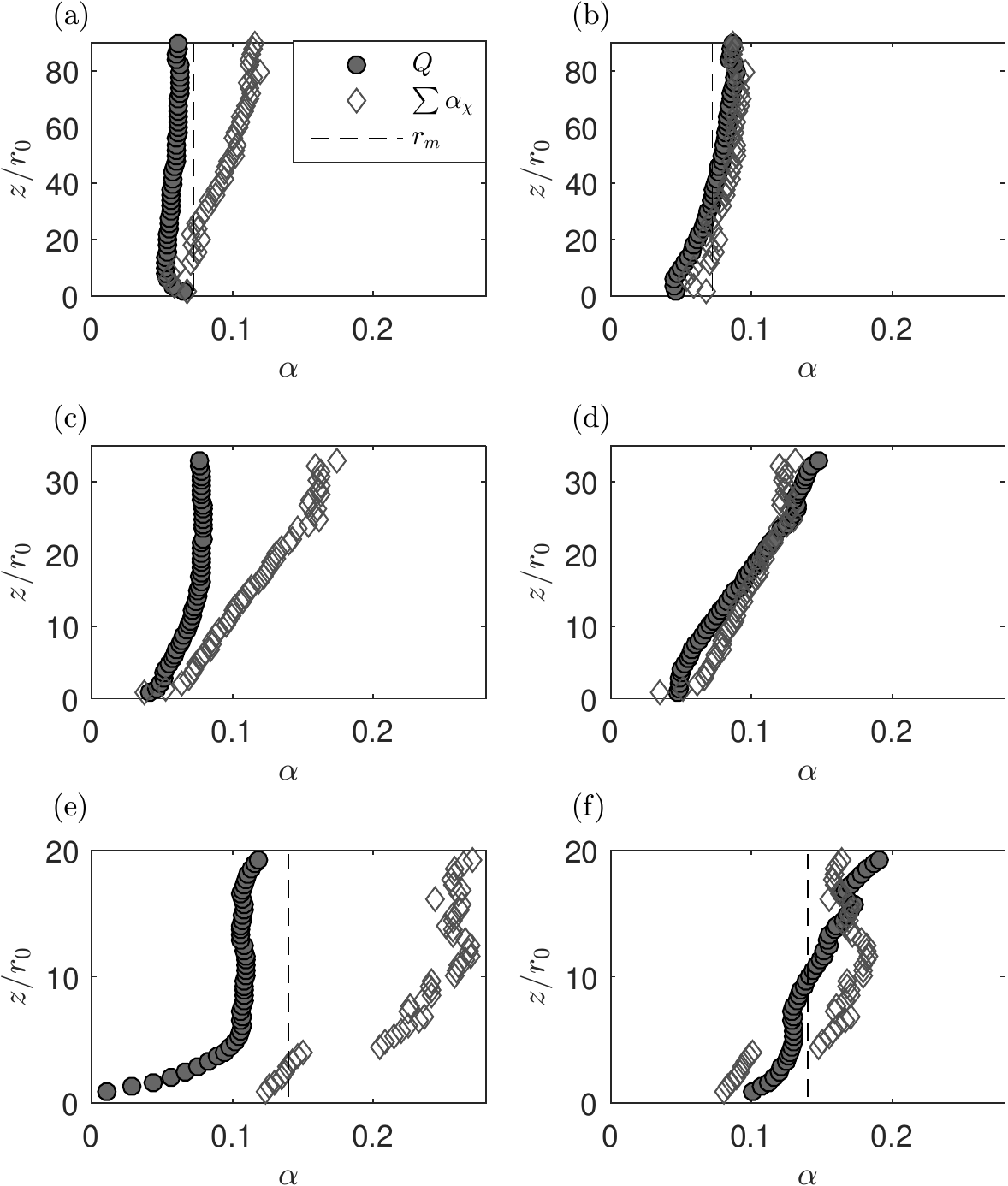}

\caption{Variation of entrainment coefficient as a function of distance above
  the source. Comparison of entrainment coefficient estimated from $r_m$, the
  volume flux balance (\ref{eq:alphadef}) and from the entrainment relation
  (\ref{eq:alphaESH15}) for J (a,b), F (c,d), P (e,f). (a,c,e)
  $\Delta w = 0.15$ ms$^{-1}$. (b,d,f) $\Delta w = 0.15 (z/z_0)^{-1/3}$
  ms$^{-1}$.}
\label{fig:alpha_comp}
\end{figure}

Figs \ref{fig:alpha_comp}(a,c,e) show the J, F and P estimates for $\alpha$
from ESH15 in the current notation.  Indicated with the dashed line in Figs
\ref{fig:alpha_comp}(a,c) is an estimate for $\alpha$ inferred from $r_m(z)$
(using the relations for $r_m$ in Table\ \ref{tab:jetplumediagram}).  All
three estimates of $\alpha$ should formally provide the same value for
$\alpha$.  For the DNS data, this is clearly the case (Fig.\
\ref{fig:alpha_decomp}), but experiments are much more difficult to control,
particularly the boundary conditions.  The measurement data
show a large discrepancy between the $Q$-based estimate for $\alpha$ and the one obtained from the entrainment relation \eqref{eq:alphaESH15}.  
This difference points to a mismatch in either the momentum balance or the mean kinetic energy balance, which can be traced back to the background flow in the ambient.

In what follows we show that the differences between our estimates for $\alpha$ can be significantly reduced by using a background motion whose magnitude is progressively reduced with distance from the source. 
As a consequence of the convection above the base plate, the plumes studied developed in a weak background velocity field that we would expect to scale as $\Delta w \sim z^{-1/3}$, i.e. the plume effectively developed within a weaker plume rising from the base plate.
By applying a background correction of the form
$\Delta w = 0.15 (z/z_0)^{-1/3}$ where $z_0$ is the distance from the plate
where the ambient vertical velocity was 0.15 m/s, the three estimates of
$\alpha$ exhibit an improved agreement, as shown in Figs
\ref{fig:alpha_comp}(b,d,f);
all estimates are in reasonably good agreement with each other. 

The method by which $\delta_m$ has been calculated for the entrainment
relation data \eqref{eq:alphaESH15} 
is performed differently than in ESH15.  Indeed, upon
close inspection of the experimental radial profiles of the Reynolds stress
$\overline{u' w'}$, in ESH15 the gradient diffusion hypothesis led to a
systematic overestimation of $\delta_m$.  As in ESH15, the $\overline{u'w'}$ profile is fitted to a function of the form
\begin{equation}
\label{eqn:F_nu_t}
\frac{\overline{u'w'}}{w_g^2} =  2 \langle\widehat{\nu_T}\rangle \frac{r}{r_g} \exp{\left(-\frac{r^2}{r_g^2}\right)},
\end{equation}
which follows from the substituting the Gaussian velocity profile
(\ref{eqn:gauss}) into the gradient-diffusion hypothesis \eqref{eq:GDH} using a constant (in $r$) eddy viscosity $\langle\nu_T \rangle = w_g r_g \langle\widehat{\nu_T}\rangle$.
However, we now consider ${r_g} $ as a free parameter (not
necessarily fixed by the value provided by the fit of (\ref{eqn:gauss})), and calculate $\langle\widehat{\nu_T} \rangle$ based on the value of $r_g$ for which the least-squares error between the measurements and \eqref{eqn:F_nu_t} is minimised.

\begin{figure}
\centering
\includegraphics{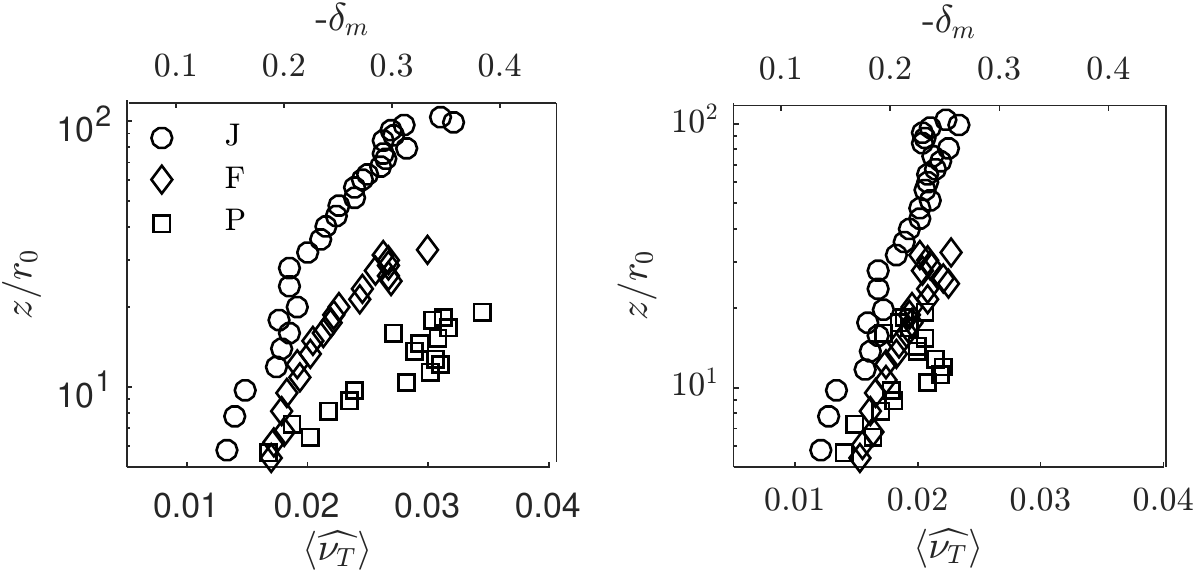}
\caption{Vertical evolution of $\delta_m$ for J, F and P.  (a) $\Delta w = 0.15$ ms$^{-1}$. (b) $\Delta w = 0.15 (z/z_0)^{-1/3}$ ms$^{-1}$.}
\label{fig:nu_t_bulk}
\end{figure}

By substituting \eqref{eqn:gauss}, \eqref{eqn:F_nu_t} into the definition for $\delta_m$, it immediately follows that $\delta_m = -8 \sqrt 2\langle \widehat{\nu_T}\rangle$; the corrected values for both $\langle \widehat{\nu_T}\rangle$ and $\delta_m$ are shown in Fig. \ref{fig:nu_t_bulk}(b).  
For all three releases, the values for $\delta_m$ are now reasonably consistent, although there is a clear increasing trend with $z$ that is not consistent with fully self-similar behaviour (in which case $\delta_m$ is expected to be constant).  Nevertheless, the data is much more consistent than the original ambient-flow correction estimate shown in Fig.\ \ref{fig:nu_t_bulk}(a).  
The data shown in Figs \ref{fig:alpha_comp}(b,d,f) and \ref{fig:nu_t_bulk}(b) was 
used to provide the input to Table 3 in \citet{vanReeuwijk2015}.

\end{document}